\documentclass[twocolumn, twocolappendix]{aastex63}

\usepackage{graphicx}
\usepackage[abs]{overpic}

\usepackage{chngcntr}

\usepackage{amsmath}
\usepackage{gensymb}

\usepackage[inline,shortlabels]{enumitem}

\usepackage{xspace}

\let\orgautoref\autoref
\providecommand{\Autoref}
{\def\equationautorefname{Equation}%
\def\figureautorefname{Figure}%
\def\subfigureautorefname{Figure}%
\def\chapterautorefname{Chapter}%
\def\sectionautorefname{Section}%
\def\subsectionautorefname{Section}%
\def\subsubsectionautorefname{Section}%
\def\Itemautorefname{Item}%
\def\tableautorefname{Table}%
\def\appendixautorefname{Appendix}%
\orgautoref}

\renewcommand{\autoref}
{\def\equationautorefname{Eq.}%
\def\figureautorefname{Fig.}%
\def\subfigureautorefname{Fig.}%
\def\chapterautorefname{Ch.}%
\def\sectionautorefname{Sect.}%
\def\subsectionautorefname{Sect.}%
\def\subsubsectionautorefname{Sect.}%
\def\Itemautorefname{item}%
\def\tableautorefname{Table}%
\orgautoref}

\newcommand{\eg}{e.g.\/,\xspace}

\newcommand{\insitu}{{in-situ}\xspace}

\newcommand{\etc}{etc.\/\xspace}

\newcommand{\Sim}{\sim\kern-0.2em\xspace}

\newcommand{\mrm}[1]{\ensuremath{\text{#1}}}
\newcommand{\txt}[1]{\text{#1}}
\newcommand{\tit}[1]{\textit{#1}}

\newcommand{\powA}[2]{\ensuremath{{{#1}\cdot10^{{#2}}}}\xspace}
\newcommand{\powB}[1]{\ensuremath{{10^{{#1}}}}\xspace}  

\newcommand{\gamlib}{\textit{GammaLib}\xspace}
\newcommand{\ctools}{\textit{ctools}\xspace}
\newcommand{\prodIrf}{\textit{prod3b-v1}\xspace}
\newcommand{\tensorflow}{\textit{tensorflow}\xspace}
\newcommand{\sncosmo}{\textit{SNCosmo}\xspace}

\newcommand{\gev}{\ensuremath{\txt{GeV}}\xspace}
\newcommand{\tev}{\ensuremath{\text{TeV}}\xspace}
\newcommand{\pev}{\ensuremath{\txt{PeV}}\xspace}

\newcommand{\erg}{\ensuremath{\txt{erg}}\xspace}
\newcommand{\scnd}{\ensuremath{\txt{s}}\xspace}
\newcommand{\yr}{\ensuremath{\txt{yr}}\xspace}
\newcommand{\hr}{\ensuremath{\txt{hr}}\xspace}

\newcommand{\pval}{\ensuremath{p}\xspace}
\newcommand{\ra}{\ensuremath{\txt{R.A.\/}}\xspace}
\newcommand{\dec}{\ensuremath{\txt{Dec.\/}}\xspace}

\newcommand{\dgr}{\ensuremath{\degree}\xspace}

\newcommand{\fermil}{{\textit{Fermi}-LAT}\xspace}

\newcommand{\IceCube}{{IceCube}\xspace}
\newcommand{\icecube}{{IceCube}\xspace}

\newcommand{\antares}{{ANTARES}\xspace}
\newcommand{\lsst}{{LSST}\xspace}
\newcommand{\plastic}{{PLAsTiCC}\xspace}

\newcommand{\cta}{{CTA}\xspace}

\newcommand{\pdf}{PDF\xspace}
\newcommand{\pdfs}{PDFs\xspace}

\newcommand{\gw}{{{GW}}\xspace}
\newcommand{\gws}{{{GWs}}\xspace}

\newcommand{\ucrs}{{{UHECRs}}\xspace}
\newcommand{\grb}{{{GRB}}\xspace}
\newcommand{\grbs}{{{GRBs}}\xspace}
\newcommand{\llgrb}{{{LL-GRB}}\xspace}
\newcommand{\llgrbs}{{{LL-GRBs}}\xspace}

\newcommand{\exgal}{{{extragalactic}}\xspace}

\newcommand{\ebl}{{{EBL}}\xspace}
\newcommand{\liso}{\ensuremath{{L_{\gamma,\txt{iso}}}}\xspace}

\newcommand{\gamray}{\ensuremath{\gamma}{-ray}\xspace}
\newcommand{\gamrays}{\ensuremath{\gamma}{-rays}\xspace}

\newcommand{\roi}{RoI\xspace}
\newcommand{\rois}{RoIs\xspace}

\newcommand{\irfs}{IRFs\xspace}

\renewcommand{\pl}{{PL}\xspace}
\newcommand{\pls}{PLs\xspace}
\newcommand{\ecut}{\ensuremath{E_{\txt{cut}}}\xspace}

\newcommand{\lstm}{LSTM\xspace}
\newcommand{\lstms}{LSTMs\xspace}
\newcommand{\ann}{ANN\xspace}
\newcommand{\anns}{ANNs\xspace}
\newcommand{\rnn}{RNN\xspace}
\newcommand{\rnns}{RNNs\xspace}

\newcommand{\dl}{DL\xspace}
\newcommand{\andet}{anomaly detection\xspace}
\newcommand{\Andet}{Anomaly detection\xspace}
\newcommand{\cls}{classification\xspace}

\newcommand{\mms}{multi-messenger\xspace}
\newcommand{\Mms}{Multi-messenger\xspace}
\newcommand{\mwl}{multiwavelength\xspace}

\newcommand{\cray}{CR\xspace}
\newcommand{\crays}{CRs\xspace}
\newcommand{\sn}{SN\xspace}
\newcommand{\sne}{SNe\xspace}

\newcommand{\ccsne}{CC-SNe\xspace}

\newcommand{\ecr}{\ensuremath{\mathcal{E}_{\mrm{CR}}}\xspace}
\newcommand{\fjet}{\ensuremath{f_{\mrm{jets}}}\xspace}

\newcommand{\lambPois}{\ensuremath{\lambda}\xspace}

\newcommand{\ts}{\text{TS}\xspace}

\newcommand{\taurnn}{\ensuremath{\tau_{\mrm{RNN}}}\xspace}
\newcommand{\tauenc}{\ensuremath{\tau_{\mrm{enc}}}\xspace}
\newcommand{\taudec}{\ensuremath{\tau_{\mrm{dec}}}\xspace}
\newcommand{\zetadec}{\ensuremath{\zeta_{\mrm{dec}}}\xspace}

\newcommand{\predClsB}{\ensuremath{P_{\text{CLAS}}^{\mrm{B}}}\xspace}
\newcommand{\predClsS}{\ensuremath{P_{\text{CLAS}}^{\mrm{S}}}\xspace}

\newcommand{\neutF}{\ensuremath{\mathcal{F}_{\nu_{\mu}}}\xspace}

\newcommand{\pdet}{\ensuremath{f_{5\sigma}}\xspace}

\newcommand{\lcdm}{\ensuremath{\Lambda}CDM\xspace}

\newcommand{\stackSignifHigh}{\ensuremath{\text{5}\sigma}\xspace}
\newcommand{\stackSignifLow}{\ensuremath{\text{3}\sigma}\xspace}

\newcommand{\subsec}{\subsection}

\shorttitle{Data-driven detection of transients}
\shortauthors{I. Sadeh}

\begin{document}

  \title{Data-driven detection of multi-messenger transients}

  \correspondingauthor{Iftach Sadeh}
  \email{iftach.sadeh@desy.de}

  \author[0000-0003-1387-8915]{Iftach Sadeh}
  \affiliation{DESY, Platanenallee 6, D-15738 Zeuthen, Germany}


  %
  \begin{abstract}

    The primary challenge in the study of explosive astrophysical transients is their detection and characterisation using multiple messengers. For this purpose, we have developed a new data-driven discovery framework, based on deep learning. We demonstrate its use for searches involving neutrinos, optical supernovae, and gamma rays. We show that we can match or substantially improve upon the performance of state-of-the-art techniques, while significantly minimising the dependence on modelling and on instrument characterisation. Particularly, our approach is intended for near- and real-time analyses, which are essential for effective follow-up of detections. Our algorithm is designed to combine a range of instruments and types of input data, representing different messengers, physical regimes, and temporal scales. The methodology is optimised for agnostic searches of unexpected phenomena, and has the potential to substantially enhance their discovery prospects.

  \end{abstract}


    %
    \section{Introduction}

        \Mms astronomy explores the universe by studying phenomena
        involving the
        electromagnetic, weak, strong and gravitational forces,
        utilising a variety of instruments~\citep{Huerta:2019rtg, Meszaros:2019xej}.
        Successful campaigns hinge on speedy and comprehensive
        coordination of observations. Examples include association of
        gravitational waves (\gws)
        from the binary neutron star merger, ${\text{GW170817}}$,
        with a short \gamray burst (\grb; \cite{2017ApJ...848L..13A}),
        as well as evidence for neutrino emission from
        the flaring blazar, ${\text{TXS\,0506+056}}$~\citep{IceCube:2018cha}.
        The main observational strategies are:
        \begin{enumerate*}[label=(\roman*),leftmargin=12pt,rightmargin=0pt,itemsep=1pt,topsep=0pt]
          \item real-time detection of signals in multiple channels;
          \item near- and late-time follow-up for direct association of events;
          \item archival stacking/population studies;
          \item correlation of multiple low-significance observables, which
          combined may result in meaningful detections.
        \end{enumerate*}
        \hspace{5pt}
        Such cross-domain analyses must reconcile
        differences in instrument sensitivities, 
        spatial and temporal coverage,
        and measurement uncertainties.
        In the following we present a new data-driven
        deep learning (\dl) framework, designed
        to tackle these challenges.

        Machine learning is a computational technique, where
        learning from examples takes the place
        of explicit functional modelling.
        %
        \dl is a type of machine learning,
        based on artificial neural networks
        (\anns; \cite{deepLearningReview, Goodfellow-et-al-2016}).
        \anns are computational models composed of \tit{neurons}, inspired
        by biological brains. Individual neurons
        perform simple transformations on vectors of inputs,
        using weight and bias parameters that are
        modified during training.
        Abstract representations of datasets can be encoded
        by arranging neurons in multiple interconnected layers,
        using nonlinear \tit{activation functions}.
        \dl utilises deep and wide layouts
        of \ann layers, able to
        represent complex models, avoiding the need for
        explicit feature engineering by domain experts.
        Effective training of these large architectures
        has become achievable due to advances in optimisation algorithms
        and in computational 
        resources.
        A prominent type of \dl is the 
        recurrent neural network (\rnn).
        Unlike \tit{feedforward} \anns, where information passes
        through a network in one direction, \rnns include
        cyclic connections; this allows them to effectively 
        process sequential data, such as time series.
        A variant
        of \rnns called a long short-term memory network (\lstm)
        has the ability to simultaneously retain information
        spanning different time scales.
        Such networks have proven very successful for, \eg
        speech recognition and natural language 
        translation~\citep{Graves:2014:TES:3044805.3045089, Sutskever:2014:SSL:2969033.2969173}.

        Deep learning has been used for a variety
        of astronomical analyses
        (see \eg~\cite{Carleo:2019ptp, Muthukrishna:2019wgc} and citations therein).
        The nominal approach has been
        to perform \tit{supervised learning}, based on labelled data.
        Examples include \gw waveforms~\citep{George:2016hay, George:2017pmj},
        neutrino detector data~\citep{2018arXiv180906166C},
        and optical images~\citep{Khan:2018opv}.
        When used for source detection,
        the inferred rate of \tit{false positives} of these methods
        strongly depends on the \tit{completeness of the training
        data}. For instance, one must verify that
        training includes all possible sources
        of systematics, including \tit{glitches} 
        (non-Gaussian noise; \cite{Wei:2019zlc, 2018PhRvD..97j1501G, Zevin:2016qwy}).
        One must also take care that such
        systematics are
        distributed in realistic proportion
        to each other and to true signal events.
        In principle, reliability of detection may be improved by
        imposing additional constraints (\eg coincidence between
        detectors, as introduced by~\cite{George:2016hay}).
        However, it remains challenging
        to directly interpret classification
        outputs as detection probabilities~\citep{Gebhard:2019ldz}.

        An alternative strategy is to use
        data-driven \tit{\andet}. The
        latter is the task of identifying data that
        differ in some respect from a reference 
        sample~\citep{PIMENTEL2014215}.
        \Andet has been used in the past in different contexts. 
        For instance, \lstms
        have been utilised to detect hardware failure
        in medical and industrial datasets,
        assuming Gaussian anomaly distributions~\citep{Malhotra2015LongST}.
        For astronomy, traditional learning methods
        have mostly been employed,
        such as principal component analysis~\citep{2019ApJ...880L..22W}
        or \tit{isolation forests}~\citep{2019MNRAS.489.3591P}.

        We present a novel use
        of \andet for the discovery and
        characterisation of astrophysical transients, utilising
        \lstm--\rnns. 
        The networks are
        coupled to a statistical pipeline used to
        interpret the results.
        Our framework facilitates combination of datasets
        of various types. It therefore
        enables derivation of realistic joint
        (\mms) probability distributions. These
        may be used for discovery, or for setting limits
        in cases of non-detection.
        Training samples may nominally be derived from an
        experiment \insitu, or from historical data.
        Compared to existing approaches, this circumvents 
        the potential pitfall of
        relying on unrepresentative reference datasets.
        Our method
        mitigates uncertainties on instrumental modelling
        and on physical backgrounds (\eg galactic foregrounds).
        It also avoids the need for 
        explicit characterisation of observing conditions,
        or of artefacts (\eg stars in the field of view).
        Finally, this data-driven approach allows for
        agnostic searches, as minimal assumptions need
        to be taken on the properties of putative sources.

        We illustrate our algorithm for three analyses,
        which represent
        the primary detection strategies of \mms astronomy:
        real-time source detection, population studies,
        and the correlation of (low-significance) observables.
        Specifically, we present examples for
        the study of high-energy transients, as
        observed with neutrinos, optical data, and \gamrays.
        
    %
    \section{Network architecture and inference pipeline}

\begin{figure*}[p]
          \begin{minipage}[c]{1\textwidth}
            \vspace{20pt}
            \begin{minipage}[c]{1\textwidth}
              \begin{center}
                \begin{overpic}[trim=13mm 105mm 53mm 14mm,clip,width=1\textwidth]{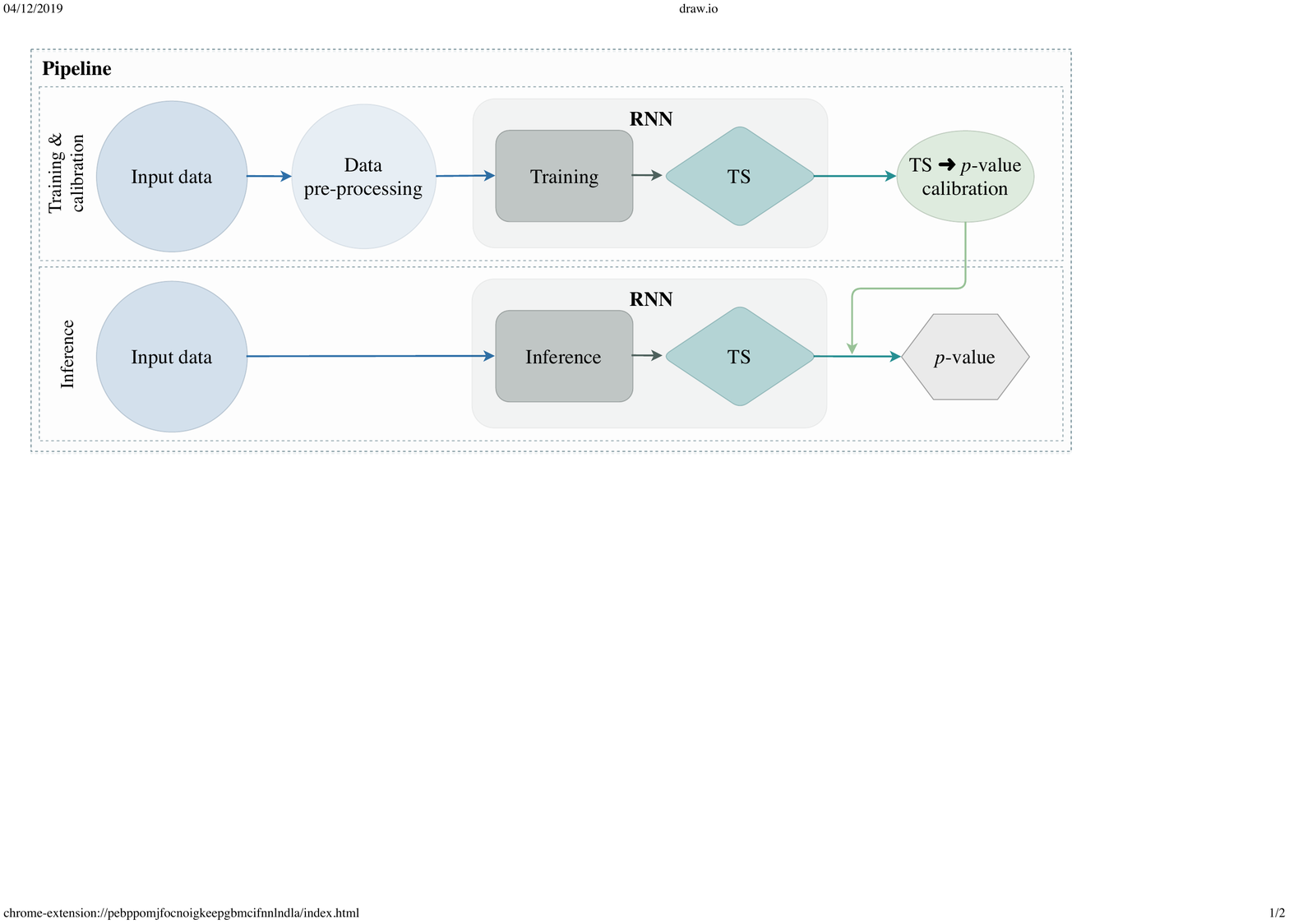}
                 \put(5,205){\textbf{A}}
                \end{overpic}
              \end{center}
            \end{minipage}\hfill
            %
            %
            \begin{minipage}[c]{1\textwidth}
              \vspace{-5pt}
              \begin{center}
                \begin{overpic}[trim=13mm 82mm 53mm 16mm,clip,width=1\textwidth]{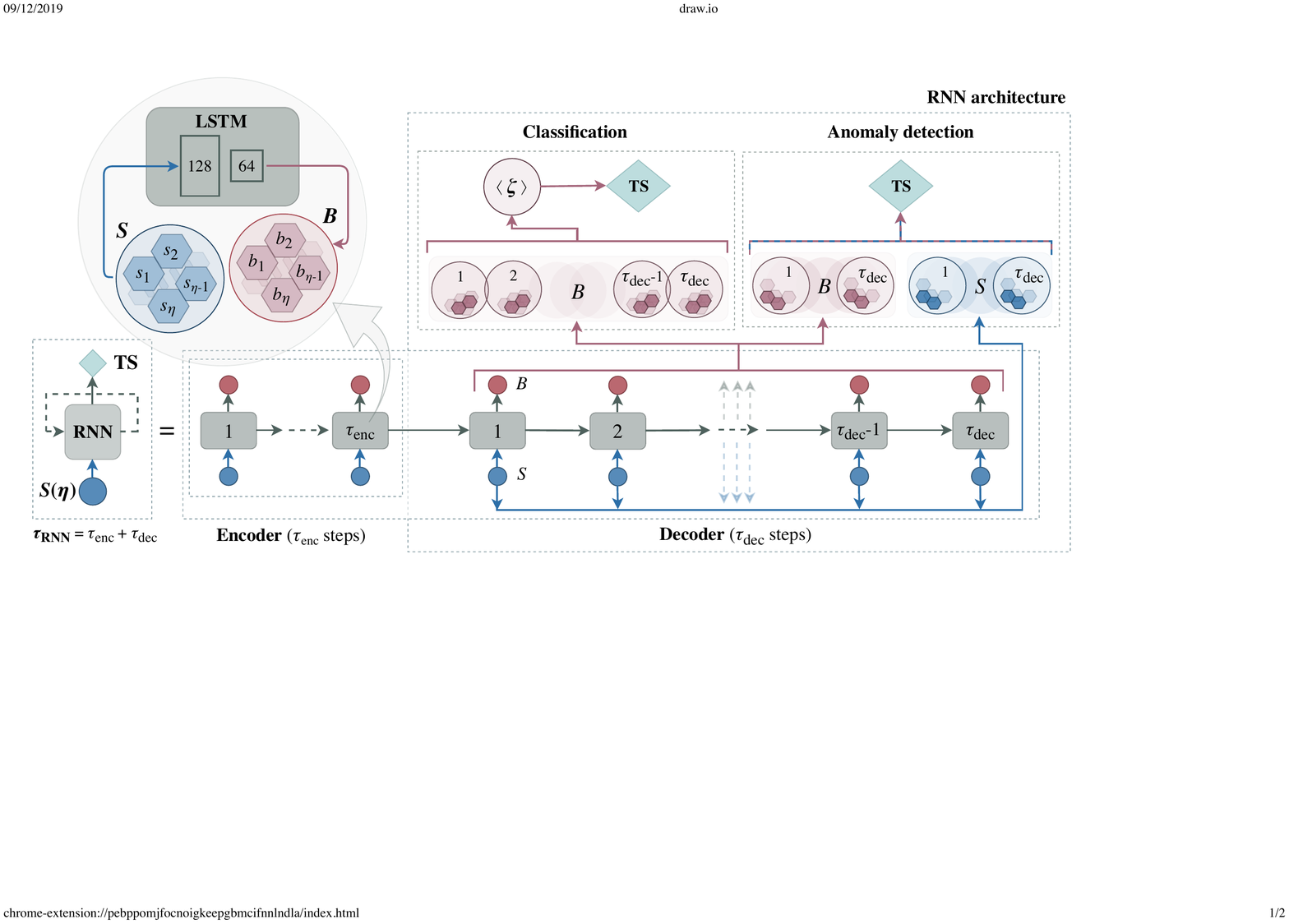}
                 \put(5,225){\textbf{B}}
                \end{overpic}
              \end{center}
            \end{minipage}\hfill
            %
            %
            \begin{minipage}[c]{1\textwidth}
              \vspace{-5pt}
              \begin{center}
                \begin{overpic}[trim=13mm 135mm 53mm 55mm,clip,width=1\textwidth]{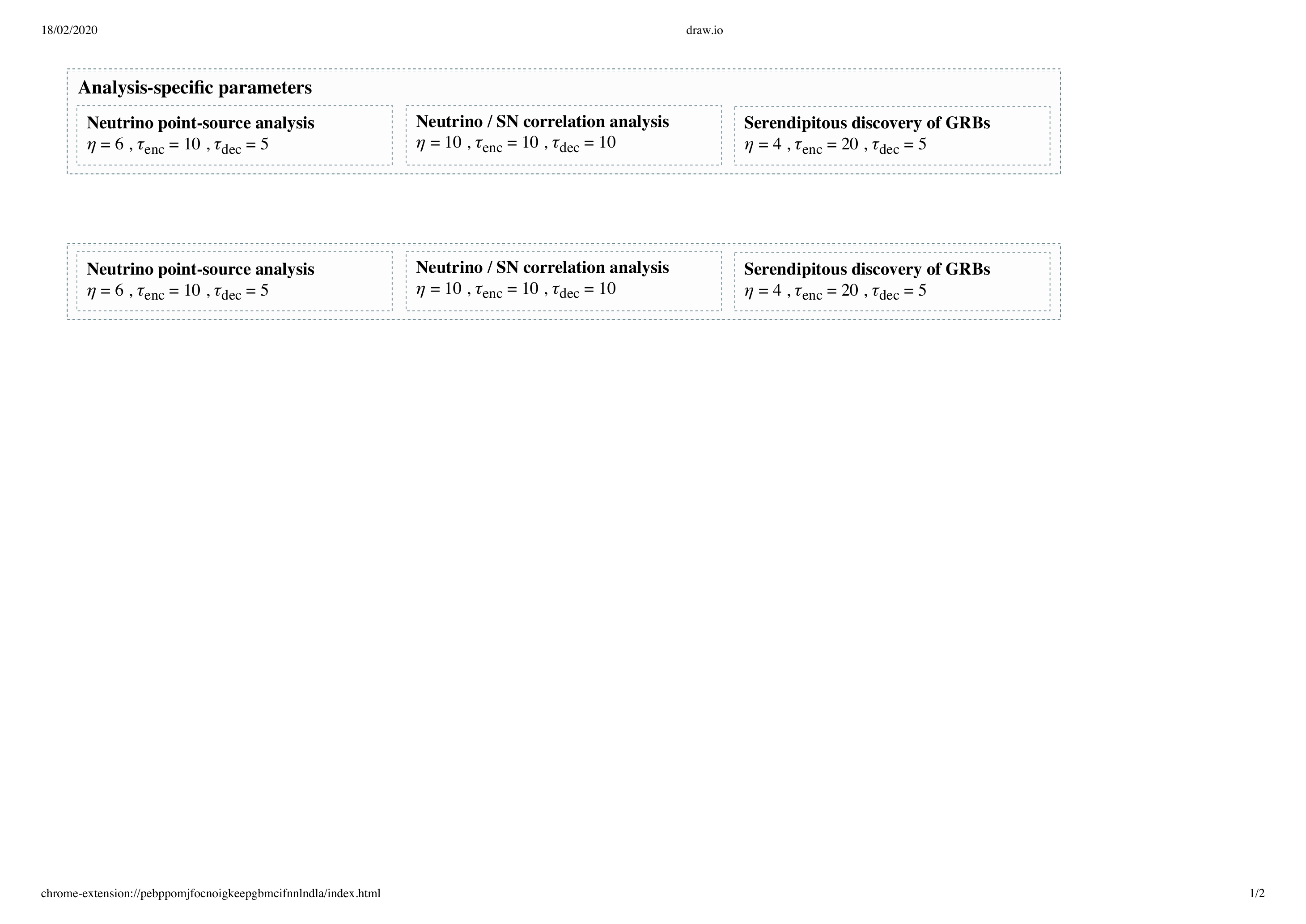}
                \end{overpic}
              \end{center}
            \end{minipage}\hfill
            %
            %
            %
            \begin{minipage}[c]{1\textwidth}
              \begin{center}
                \begin{minipage}[t]{1\textwidth}\begin{center}
                  \caption{\label{FIGannArch}\textbf{Schematic
                    illustrations of the software pipeline,
                    and of the architecture of the recurrent 
                    neural network (\rnn).}
                    \textbf{(A)}~The
                    pipeline comprises two main phases,
                    \tit{training/calibration} and \tit{inference}.
                    Training includes a pre-processing stage
                    for generation of background simulations.
                    %
                    These data are used to train the \rnn
                    and calculate test statistics (\ts). 
                    The latter are mapped to \pval-values as part
                    of the \tit{calibration} phase.
                    %
                    Inference includes evaluation of \ts- and \pval-values,
                    using the trained \rnn and the
                    pre-calculated calibration.
                    \textbf{(B)}~The network may be decomposed into
                    an \tit{encoder} (\tauenc time-steps)
                    and a \tit{decoder} (\taudec time-steps),
                    where the decoder represents the search interval.
                    %
                    The \rnn is made up of long 
                    short-term memory networks (\lstms)
                    (green rectangles).
                    Each \lstm comprises two layers
                    of~128 and 64~hidden units.
                    The input data, $S(\taudec, \eta)$
                    (blue circles),
                    make up ${\eta}$ numbers for each time
                    step (blue hexagons).
                    Similarly, the outputs of the \lstms are
                    indicated as $B(\taudec, \eta)$ (red circles).
                    %
                    For \tit{\andet}, the decoder inputs and 
                    outputs, $S$ and $B$, 
                    are directly used to calculate the \ts
                    for discovery. 
                    For \tit{\cls}, the decoder outputs are fed into
                    logits (proxies for classification probabilities),
                    ${\zeta(\taudec)}$,
                    which are used to
                    derive the corresponding \ts.
                    The particular set of parameters used for
                    each one of the example analyses
                    ($\eta$, \tauenc, and \taudec) are
                    are shown in the bottom panel. The specific
                    choices are described in detail in the text.}
                \end{center}\end{minipage}\hfill
              \end{center}
            \end{minipage}\hfill
            \vspace{15pt}
          \end{minipage}\hfill
        \end{figure*}

        Our software pipeline and chosen \rnn
        architecture are shown in \Autoref{FIGannArch}.
        The \rnn may be decomposed into two
        elements, an \tit{encoder} and a
        \tit{decoder}~\citep{2014arXiv1406.1078C}.
        The \rnn accepts ${\taurnn = \tauenc + \taudec}$~time
        steps as input.
        The first \tauenc~encoder steps
        represent the background
        interval, before a transient appears.
        A potential signal event is
        searched for within the
        following ${\taudec}$~decoder steps.
        For each of the example analyses described below,
        the value of \taurnn is chosen according
        to the expected properties of signals
        (\eg physical time-scales),
        accounting for the structure of the available data.
        Each step receives a collection of
        (analysis-specific) ${\eta}$~inputs.

        In our nominal approach we employ an
        \andet technique. 
        We utilise sequences of input data, $S(\taurnn, \eta)$, which
        for training correspond to the response of
        an instrument in the absence of signal events. 
        The \rnn
        is used to predict the
        expected background of the experiment, $B(\taudec, \eta)$.
        Transients are then detected as significant divergences
        from these predictions.
        We define a unique test statistic (\ts),
        which encapsulates the difference (\tit{mean squared error})
        between $S$ and $B$ for each analysis.

        Additionally, we employ a complementary
        classification approach (illustrated for the \gamray
        analysis example). In this case, the \rnn is used
        to to directly classify transient
        events, rather than to predict the background.
        Correspondingly, the
        network is trained using labelled examples
        of both background and putative signal data.
        The discovery \ts is based on the 
        ratio between the background and signal classification 
        probabilities~\citep{Cranmer:2015bka, Goodfellow-et-al-2016}.

        The \ts output
        of the \rnn is coupled to a pipeline that is
        used to derive the significance of detection.
        Nominally, we do not assume a specific statistical
        model for the background or for the signal. We therefore
        generate multiple
        realisations of the input background sample,
        from which we derive cumulative distributions of \ts.
        The latter are used to calibrate the relationship between \ts-values
        and \pval-values for a background-only test hypothesis.

        We train the \rnns using \tit{Adam}
        optimisation with a
        \tit{learning rate} of~${\text{0.01}}$.
        %
        For each of the three \andet analyses,
        we have ${\text{\powB{4}}}$ background
        sequences of \taurnn steps.
        For the single \cls example,
        an additional
        ${\text{\powB{4}}}$ signal events are used as well.
        The inputs are independently normalised, such that their
        nominal range of values for the background
        training sample is mapped to the interval,~${[0,\,1]}$.
        Data are randomly split into batches
        of 64~sequences for training,
        where ${\text{20}}\%$ of events are
        set aside for validation.
        In order to mitigate over-fitting,
        we apply 30\% \tit{dropout} training regularisation
        (random masking of units).
        %
        The process of training for the various analyses
        lasts several minutes
        on a laptop--CPU, with
        corresponding sub-sec inference 
        (well below the requirements on
        the relevant real-time applications).

        We compared several configurations of 
        \rnn hyper-parameters (internal configuration
        parameters of the network).
        For instance, this included
        doubling the number
        of \lstm layers, increasing the batch size, varying the
        learning rate, \etc 
        In addition, we performed analysis-dependent systematic
        checks, as described below.
        We found no significant variation in the results.
        The robustness in performance is due to our design
        choice of a simple \rnn, and to the fact that
        we rely on data-driven calibration of test statistics
        post-training.

        In the following, we illustrate our framework
        using concrete examples.
        We restrict the discussion
        to the general features of the method in each case, such
        as relevant time-scales, \rnn inputs,
        and significance of detection.
        For a comprehensive description of the datasets, 
        definition of test statistics,
        source modelling, 
        and systematic tests,
        see \autoref{sec:appendix:1}.

    %
    \section{Results}
        \subsec{Neutrino point-source search}

            A diffuse {\tev--\pev} flux of astrophysical
            neutrinos has been discovered by \icecube~\citep{Aartsen:2013jdh}.
            While the exact nature of the emission
            remains elusive, its apparent isotropy suggests that it originates
            from relatively weak \exgal sources.
            Possible sources of ultra high-energy
            cosmic rays (\ucrs) and
            neutrinos include long- and
            short-duration 
            gamma-ray bursts (\grbs),
            as well as  core-collapse supernovae (\ccsne) with choked
            jets or shock breakouts~\citep{2013ApJ...769L...6K, Senno:2017vtd}.
            The energy density of the astrophysical neutrinos is comparable
            to that of the isotropic \gamray background and to that of
            \ucrs~\citep{Meszaros:2019xej}.
            This indicates that \mms
            interpretation may lead to breakthroughs in our
            understanding of cosmic ray (\cray) accelerators. It
            may elucidate
            the connection between supernovae (\sne) and \grbs, and
            may shed light on the nature of their central engine.
            Searches commonly take the form of either
            spatio-temporal clustering of neutrinos,
            or their direct association with steady or transient
            sources~\citep{Aartsen:2015wto}.

            We search for clustering in two all-sky samples.
            The first comprises \icecube event lists
            of track-like muon neutrino candidates, taken between
            ${\text{2011}}$ and ${\text{2012}}$ (MJD~55694--56415; \cite{Aartsen:2016oji}).
            The second is of \antares muon neutrinos, observed within
            the same time period~\citep{Adrian-Martinez:2014wzf}.
            The nominal inputs to our \rnn are based on 
            \tit{neutrino event density} metrics
            (see \autoref{sec:appendix:1:2}).
            The densities are defined
            with respect to a given location on the sky.
            For the \icecube sample, the data
            are split into four logarithmically
            spaced bins in the energy proxy,
            between ${\text{10}}$~\gev and ${\text{8}}$~\pev.
            For \antares the metrics are inclusive in energy.
            The data are integrated over
            ${\text{24}}$~\hr time
            periods, which effectively avoids dependence of
            event rates on \ra~\citep{Aartsen:2015wto}.
            The response of the \icecube and \antares
            detectors depends on the zenith
            of the observed event, which is therefore
            added as input to the \rnn. 
            In cases where an \icecube source is not
            within the field of view of \antares,
            we use background data instead.
            In total, we have ${\eta=\text{6}}$~inputs per
            time step.
            The collection of inputs is derived for 
            ${\taurnn=\text{15}}$~days periods. 
            The first ${\tauenc=\text{10}}$~days are assumed to only
            contain background events. Transient
            signals are searched for within the next ${\taudec=\text{5}}$~days
            interval. In total, the \rnn 
            receives $\text{6}\times\text{15}=\text{90}$~inputs.

            \begin{figure*}[t]
                \begin{minipage}[c]{1\textwidth}

                \begin{minipage}[c]{1\textwidth}
                \begin{center}
                    \includegraphics[trim=0mm 3mm 0mm 20mm,clip,width=.78\textwidth]{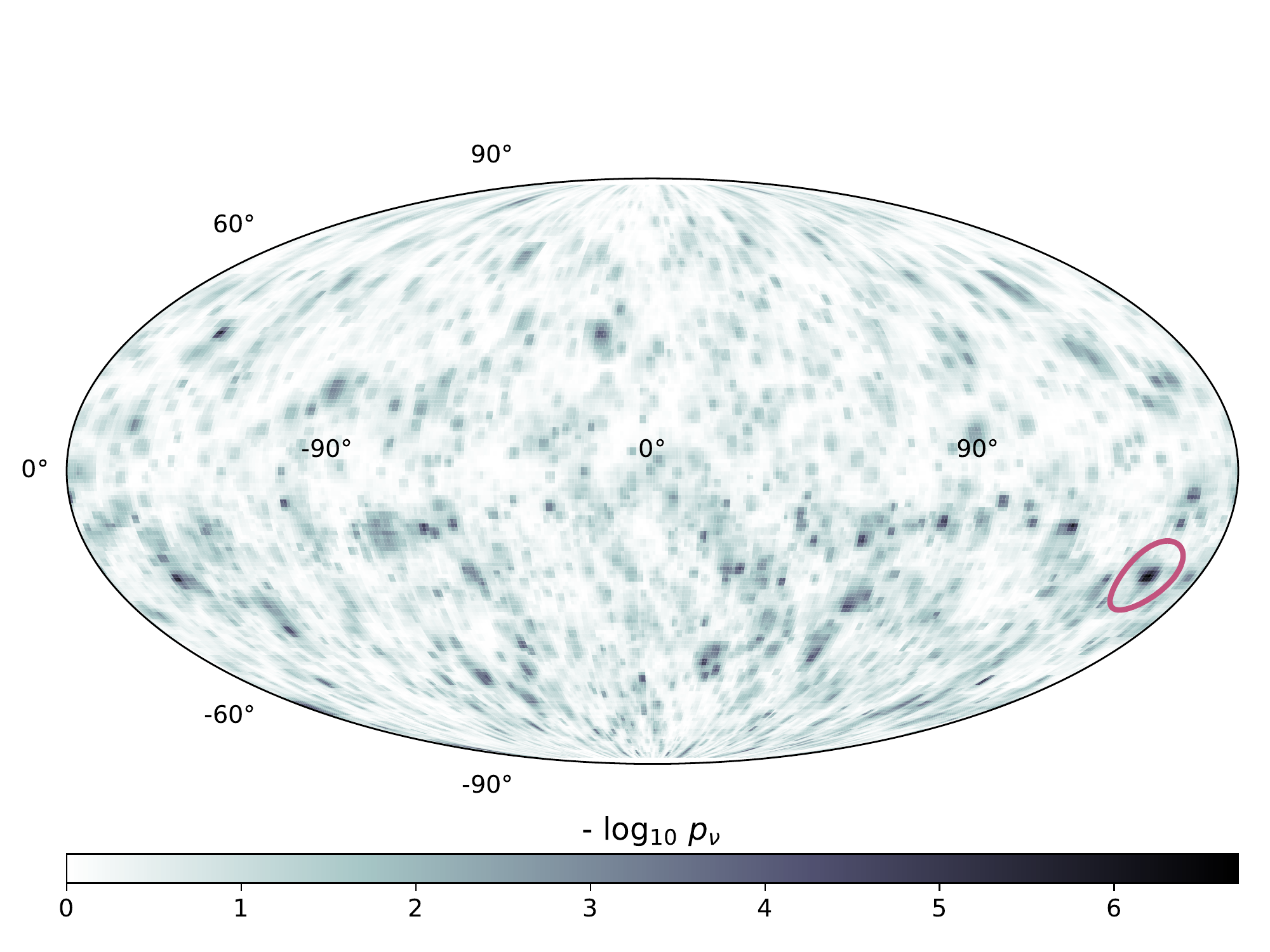}
                \end{center}
                \end{minipage}\hfill
                %
                %
                \begin{minipage}[c]{1\textwidth}
                \begin{center}
                \begin{minipage}[t]{1\textwidth}\begin{center}
                \caption{\label{FIGneutrinosPoint} \textbf{Results
                    of the neutrino point-source search analysis.} The
                    sky-map in equatorial coordinates
                    displays the pre-trials \pval-values,
                    ${p_{\nu}}$, for time-dependent
                    neutrino point sources, utilising \icecube and \antares
                    data from ${\text{2011--2012}}$.
                    The most significant point
                    ${\left( \{\ra,\,\dec\} = \{163\dgr,\,-26.5\dgr\} \right)}$
                    is indicated by the red oval. It
                    has ${p_{\nu}=\powA{1.9}{-7}}$ and 
                    ${p_{\nu}=0.1}$ pre- and post-trials, respectively,
                    corresponding to $5.2\sigma$
                    and $1.6\sigma$ significance.
                    The result does not correspond to
                    any astronomical object
                    of interest, and is consistent with
                    a statistical fluctuation.}
                \end{center}\end{minipage}\hfill
                \end{center}
                \end{minipage}\hfill
                \vspace{5pt}
                \end{minipage}\hfill
            \end{figure*} 

            The \rnn is trained to predict the
            neutrino event densities
            in each of the five days being probed, for a particular
            sky position.
            As part of \andet, these predictions are meant to correspond
            to the background. Potential
            transient signals must therefore be removed from the
            training dataset. This is done
            by scrambling neutrino events in \ra and 
            time of detection.

            The test statistics for detection are based on differences
            between the predictions of the network and the
            true data.
            For our particular definitions of observables, one can
            not assume a priori that a particular statistical model
            (\eg Poissonian counts) would be valid.
            We therefore calculate detection
            probabilities in a more general approach, using simulations.
            We consider the zenith to be an
            \tit{auxiliary \rnn parameter}, as
            it is not directly used
            to derive detection probabilities.
            Rather, we bin the data in zenith
            into~90 intervals of~${\text{2}\dgr}$ width, in which
            the detector response is approximately uniform.
            We then generate~${\Sim\powB{6}}$ scrambled background
            sequences for each bin and evaluate them with the \rnn.
            The derived distributions of test statistics are used to parametrise
            the correspondence between \ts- and \pval-values, accounting
            for all spatial and temporal trials.

            We proceed to use the trained network to evaluate the
            original (non-scrambled) data, carrying out 
            an all-sky grid search
            with~${\text{0.3}\dgr}$ spacing.
            The corresponding spatial distribution of \pval-values
            is shown in \Autoref{FIGneutrinosPoint}.
            The most significant position is
            ${\{\ra,\,\dec\} = \{163\dgr,\,-26.5\dgr\}}$. It has
            ${p_{\nu}=\powA{1.9}{-7}}$ (pre-trials) and 
            ${p_{\nu}=0.1}$ (post-trials),
            corresponding to $\text{5.2}\sigma$ and
            $\text{1.6}\sigma$ significance.
            The result does not correspond to any astronomical object
            of interest, and is consistent with a statistical fluctuation.
            We also use the outputs of the \rnn to
            perform a correlation analysis between the
            \icecube and \antares events
            (see \autoref{sec:appendix:1:2}),
            finding no significant result.

            Our conclusions on the existence
            of neutrino sources are consistent with previous
            studies of these data, which did not detect
            any source (${p_{\nu}=0.6}$, by~\cite{Aartsen:2015wto}).
            %
            In this case, however, the analysis is done
            without the need to explicitly
            define likelihood functions for the background or sources.
            The study is performed on
            a combined \icecube and \antares dataset.
            Another advantage of our approach, is that
            there is no need to model the relative response between
            the two neutrino observatories.
            This example also illustrates the flexibility of 
            the methodology regarding observables.
            Our choice of inputs and test statistics
            is motivated by the properties of the datasets, but
            is by no means unique. However, our framework
            is designed to provide self-consistent
            detection probabilities in general, given a set of
            primary (\eg neutrino densities) and auxiliary (\eg zenith)
            parameters.
            Finally, we note that
            the limitations of the public datasets
            constrain us to $\geq\text{1}$~day time bins.
            Provided that the full data-streams of the
            experiments become available,
            the same approach would be applicable for real-time
            searches on shorter time-scales.

        %
        \subsec{Correlation analysis between neutrinos \\and \ccsne}

            An alternative to auto-correlation analyses
            is to search for cross-correlation between different messengers.
            An important physical example is that of
            core-collapse supernovae
            with relativistic 
            outflows~\citep{Murase:2013ffa, Cano:2016ccp}.
            %
            \ccsne may accelerate \ucrs, which produce \gamrays
            and neutrinos,
            for instance, via~${pp}$ or~${p\gamma}$
            interactions.
            %
            \begin{figure*}[p]
                \begin{minipage}[c]{1\textwidth}
                \begin{minipage}[c]{1\textwidth}
                \begin{minipage}[c]{0.5\textwidth}
                \begin{center}
                \begin{overpic}[trim=9mm -12.5mm 0mm 60mm,clip,width=.98\textwidth]{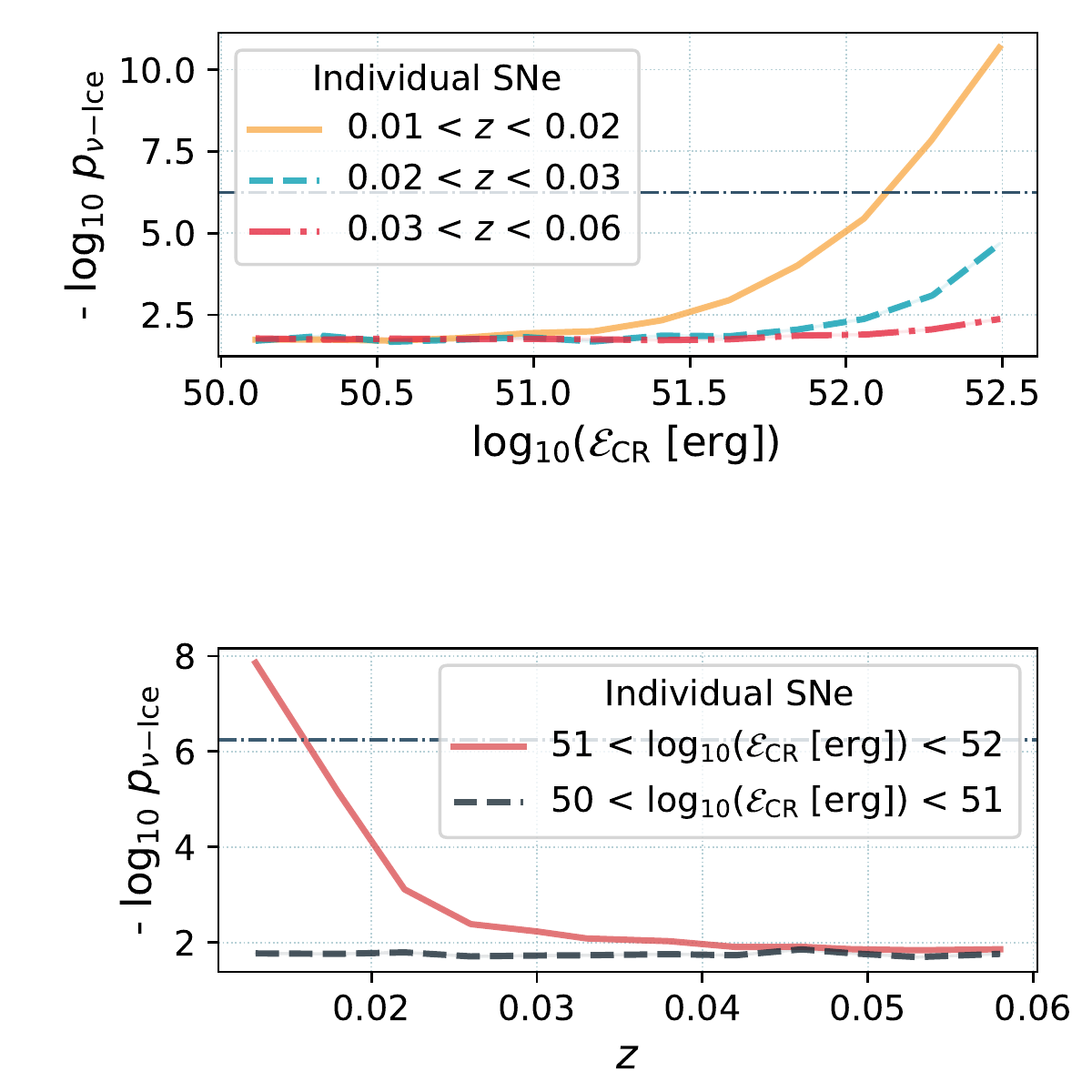}
                \put(34,145){\textbf{A}}
                \end{overpic}
                \end{center}
                \end{minipage}\hfill
                \begin{minipage}[c]{0.5\textwidth}
                \begin{center}
                \begin{overpic}[trim=7mm 66mm 2mm 1.5mm,clip,width=.98\textwidth]{figures/sn_neutrinos}
                \put(38,123){\textbf{B}}
                \end{overpic}
                \end{center}
                \end{minipage}\hfill
                \vspace{-20pt}
                \end{minipage}\hfill
                %
                %
                \begin{minipage}[c]{0.5\textwidth}
                \begin{center}
                \begin{overpic}[trim=9mm 0mm 0mm 0mm,clip,width=.98\textwidth]{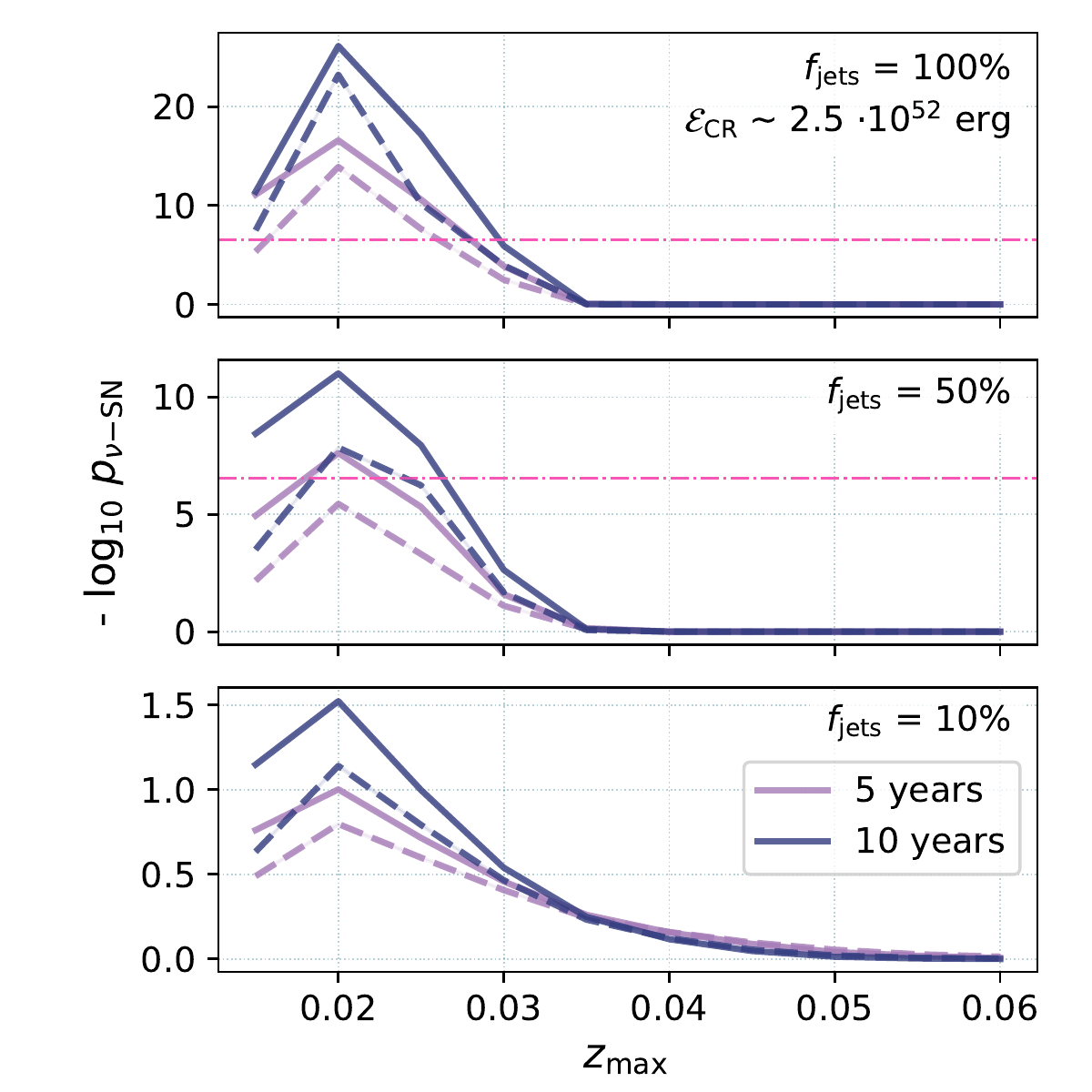}
                \put(34,270){\textbf{C}}
                \end{overpic}
                \end{center}
                \end{minipage}\hfill
                \begin{minipage}[c]{0.5\textwidth}
                \begin{center}
                \begin{overpic}[trim=7mm 0mm 2mm 0mm,clip,width=.98\textwidth]{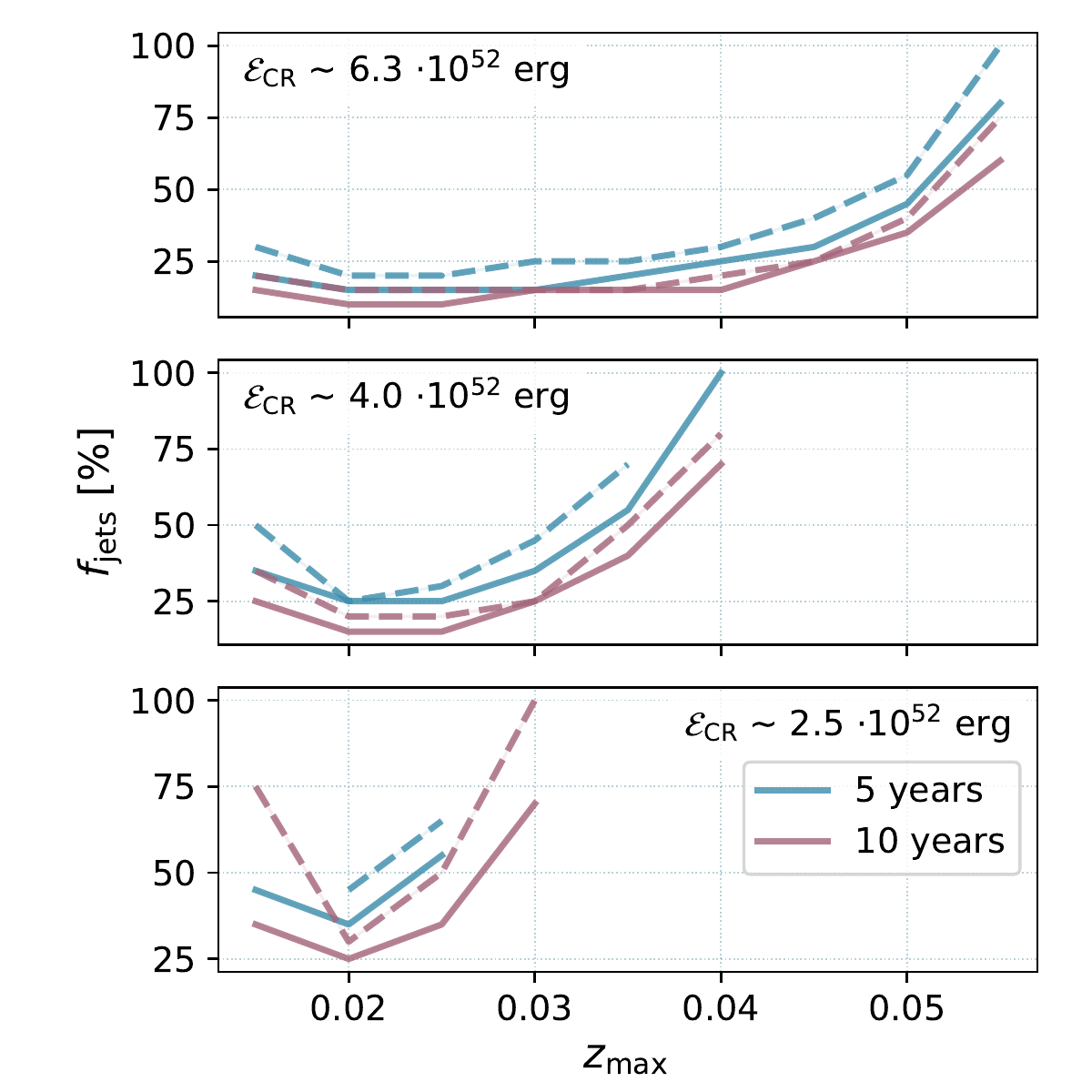}
                \put(38,270){\textbf{D}}
                \end{overpic}
                \end{center}
                \end{minipage}\hfill
                %
                %
                \begin{minipage}[c]{1\textwidth}
                \begin{center}
                \begin{minipage}[t]{1\textwidth}\begin{center}
                \caption{\label{FIGneutrinosSN} \textbf{Results of the
                  correlation analysis between neutrinos and \ccsne.}
                  $\text{{\textbf{(A)-(B):}}}$~The
                  top~${1\%}$ (most significant)
                  of the distribution of
                  \pval-values, ${p_{\nu\mrm{-Ice}}}$, for neutrinos
                  associated with individual \sne (not corrected
                  for trials),
                  \textbf{(A)}~as a function of the redshift, $z$;
                  and \textbf{(B)}~as a function
                  of the energy of a source
                  deposited into cosmic rays, \ecr.
                  The dashed--dotted horizontal lines highlight
                  the value corresponding to the pre-trials~${5\sigma}$
                  detection significance of a single event.
                  The combined values of ${p_{\nu\mrm{-Ice}}}$ from
                  different samples of \sne serve as
                  the basis for the stacking analysis.
                  %
                  %
                  \textbf{(C):}~Post-trials 
                  \pval-values, ${p_{\nu\mrm{-SN}}}$,
                  for the stacked sample of neutrinos, as a function of
                  the maximal redshift of observed \sne included
                  in the analysis, $z_{\mrm{max}}$.
                  Dashed and full lines, respectively, correspond to
                  ${\sigma_{\text{SN}}^{\text{min}} = \stackSignifHigh}$
                  and~\stackSignifLow for the
                  optical detection threshold for individual \sne.
                  The dashed--dotted horizontal line in the top panel highlights
                  the value corresponding to~${5\sigma}$ significance
                  for the stacked search.
                  As indicated, we consider $\text{5}$ and $\text{10}$~\yr~\lsst surveys,
                  where it is assumed that
                  all \sne have ${\ecr = \powA{2.5}{52}~\erg}$.
                  The three panels
                  illustrate the results for different values of \fjet,
                  the fraction of \sne for which neutrino emission
                  is observable.
                  In general, ${\fjet\ll1}$ is expected for
                  beamed emission, and ${\fjet\Sim1}$ for
                  shock breakouts.
                  \textbf{(D):}~The sensitivity
                  of the stacking analysis, expressed as the minimal
                  value of \fjet, for which~${\geq5\sigma}$
                  detection is achievable. 
                  Different values of \ecr are compared
                  in the three panels, 
                  where dashed and full lines, respectively, correspond to
                  ${\sigma_{\text{SN}}^{\text{min}} = \stackSignifHigh}$
                  and~\stackSignifLow.}
                  %
                  %
                \end{center}\end{minipage}\hfill
                \end{center}
                \end{minipage}\hfill
                \vspace{5pt}
                \end{minipage}\hfill
            \end{figure*} 
            %
            A direct connection between neutrinos and their \sn counterparts
            can be made for events that also exhibit high-energy
            emission, \eg~\grbs.
            However, \gamrays are
            not always observed in association with \ccsne.
            For example, depending on the environment within and
            around a source, \gamrays
            may become attenuated due to ${\gamma\gamma}$ interactions~\citep{2019ApJ...872..110B}.
            A few percent of \ccsne are estimated to harbour
            undetectable jets, which are not
            powerful enough to punch through their
            progenitors and winds.
            Such events are associated with choked jets, and possibly
            shock breakouts~\citep{2013ApJ...769L...6K}.
            %
            Conversely, those jets that are successful are mostly
            launched off-axis with respect to the observer, 
            and thus are also 
            undetected~\citep{Denton:2017jwk}.
            %
            Finally, even when the high-energy
            emission is beamed towards Earth,
            individual \sne are
            unlikely to be identified as sources of neutrinos,
            as illustrated in \Autoref{FIGneutrinosSN}~(A-B).
            This motivates performing a stacking analysis,
            combining observations from many events.

            %
            We simulate a neutrino stacking analysis, intended
            to identify an accumulated
            neutrino over-density 
            in spatio-temporal coincidence with 
            \ccsne~\citep{Senno:2017vtd}.
            This should be distinguished from the more direct
            cross-correlation approach, which
            involves optical follow-up of specific
            neutrino events~\citep{2019ICRC...36..963M}.
            A direct analysis is in principle preferred, as it
            enables the detailed study
            of a particular source~\citep{IceCube:2018cha}.
            However, it has two main disadvantages for
            the case of \ccsne. Firstly,
            one must have high confidence that a given neutrino is
            astrophysical, which generally constrains events to
            very-high energies. 
            Additionally, the sensitivity is limited
            by the irreducible contamination of unassociated
            \sne within the uncertainty region of
            the neutrino.
            Given the weak nature of neutrino sources
            and the large number of potential counterparts,
            indirect population studies become competitive.

            The purpose of the \rnn in the stacking analysis
            is to provide
            a background model for the joint neutrino
            and optical observation.
            Optical transients are 
            simulated according to
            projected observations with the
            Wide-Fast-Deep survey of the
            upcoming Vera C.~Rubin Observatory
            (previously referred to as the Large 
            Synoptic Survey Telescope, LSST; \cite{2009arXiv0912.0201L}).
            These \sne are
            only used to define
            time-intervals, during which neutrino flares
            may occur. 
            Neutrino signals, based on our
            \icecube sample, are integrated over these intervals.

            The inputs to the \rnn are of two types. 
            For a given time step, the
            first consists of the \icecube neutrino
            densities (as for the previous example).
            The second type consists of \lsst
            signal-to-noise metrics~\citep{2017ApJ...836..187Z} in 
            five optical bands
            (see \autoref{sec:appendix:1:3}).
            The zenith of observation is
            also added as an auxiliary parameter.
            We thus have ${\eta=\text{10}}$~inputs per
            time step in total.
            \sne light curves evolve over days--weeks.
            We therefore construct 
            ${\taurnn=\text{20}}$~days data sequences, 
            with the first ${\tauenc=\text{10}}$~days representing
            the background, and the 
            next~${\taudec=\text{10}}$ the search period.
            In total the \rnn 
            receives $\text{10}\times\text{20}=\text{200}$~inputs.

            We begin by creating background samples. These represent
            periods of joint neutrino and optical observations,
            in which no transient exists in either messenger.
            The neutrino data are derived by the
            scrambling procedure described above.
            The optical data are chosen based on the 
            true peak time of simulated \sne.
            The background samples are used to train the \rnn.
            For a particular sky position, we then derive
            individual test statistics for source detection
            with each messenger, which are calibrated
            to \pval-values as a function
            of zenith.

            In the next step of the analysis, we simulate
            stacked neutrino signal samples.
            For the source model, we
            assume that optically detected 
            \sne are in fact \grbs, and that neutrinos
            are emitted during the short prompt phase of explosions. 
            The fluence of muon neutrinos is modelled as
            $\neutF \propto \ecr / D^{2}_{\mrm{L}}$,
            where \ecr is the energy deposited in \crays,
            and ${D_{\mrm{L}}}$ is the luminosity distance
            to the source~\citep{Waxman:1998yy,Senno:2017vtd}.
            We consider different values for \fjet, 
            the fraction of \sne for which the neutrino emission
            is directed towards the Earth.
            Depending on the physical model, the
            emission may \eg be collimated
            (relativistic jets; ${\fjet \ll1}$), or 
            quasi-spherical (shock breakouts; ${\fjet \Sim 1}$).
            We also account for \sn misclassifications,
            in which cases neutrino signals are not injected.
            For this simplified example, we assume that
            \ccsne may be identified from their photometric light curves
            with a ${\text{10}\%}$ fake-rate~\citep{Muthukrishna:2019wgc}.

            The \ts of the stacked signal is computed by
            averaging the individual neutrino
            detection significance of all events passing
            a selection threshold, ${\sigma_{\text{SN}}^{\text{min}}}$.
            This threshold is defined
            by the detection significance of the corresponding 
            optical \sn.
            In order to derive the final stacked detection significance,
            we simulate the background hypothesis;
            we impose the nominal optical 
            detection procedure, but do not
            inject any signals into the neutrino data.
            We generate~${\Sim\powB{7}}$ such
            realisations of a full \lsst survey, accounting for
            random coincidence between observables.

            The results of the analysis are shown in \Autoref{FIGneutrinosSN}.
            We compare different values of \fjet
            and \ecr, different optical detection
            thresholds (${\sigma_{\text{SN}}^{\text{min}} = \stackSignifLow,\,\stackSignifHigh}$),
            and different durations of \lsst surveys (${\text{5,\,10}}$~\yr).
            The search would be sensitive in the case of high
            \fjet and \ecr values, in accordance with~\cite{Senno:2017vtd}.
            Using our method, it would be possible to
            \eg significantly constrain
            shock breakout scenarios.
            If no signal is detected, the sensitivity curves could
            be used to derive physical limits on neutrino emission.

            For the neutrino point-source search,
            the \rnn was used to derive joint detection probabilities
            for two experiments. Similarly
            in this case, we avoid having to
            impose a specific relationship between
            two messengers.
            In addition, our
            approach provides a statistical framework to 
            optimise searches, and to derive limits on non-detections.
            In this illustrative example,
            we use the \rnn to tune the redshift range 
            and to relax the detection threshold
            of individual events.
            This circumvents the need to a priori refine
            the selection criteria of \sne.
            While we do not explicitly
            incorporate \sn classification or redshift estimation
            as part of our pipeline,
            such extensions are feasible~\citep{Muthukrishna:2019wgc}.
            They are planned for future work, paving the way for
            real-time applications.

        %
        \subsec{Serendipitous discovery of \grbs}

            For \sne engines of sufficient power, relativistic
            jets manage to break out of the progenitor.
            Depending on their inclination, these
            may be observed as long \grbs.
            %
            %
            \begin{figure*}[t]
                \begin{minipage}[c]{1\textwidth}
                \vspace{5pt}
                \begin{minipage}[c]{0.5\textwidth}
                \begin{center}
                \begin{overpic}[trim=0mm 12mm 0mm 13mm,clip,width=.98\textwidth]{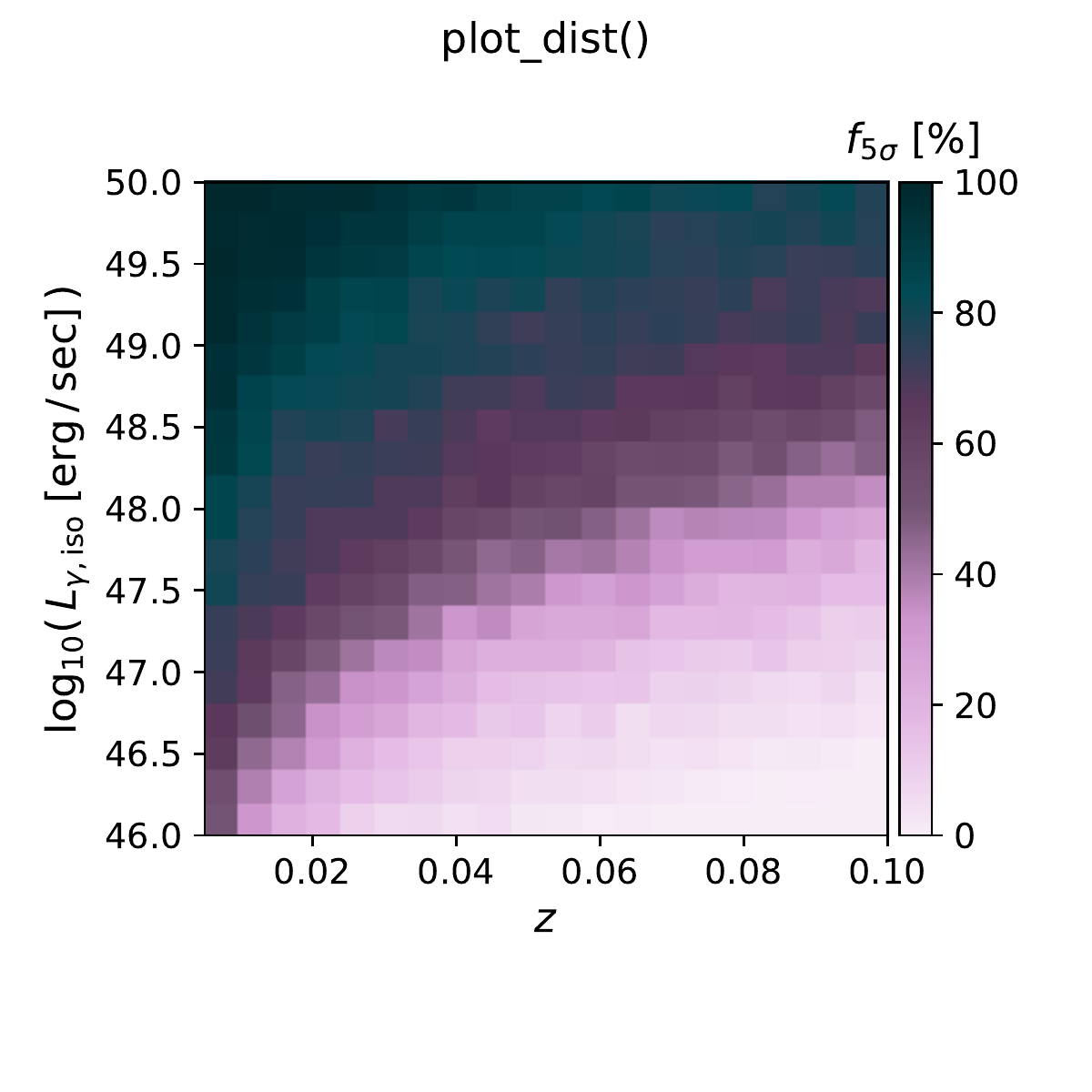}
                \put(48,193){\textbf{A}}
                \end{overpic}
                \end{center}
                \end{minipage}\hfill
                \begin{minipage}[c]{0.5\textwidth}
                \begin{center}
                \begin{overpic}[trim=0mm 12mm 0mm 13mm,clip,width=.98\textwidth]{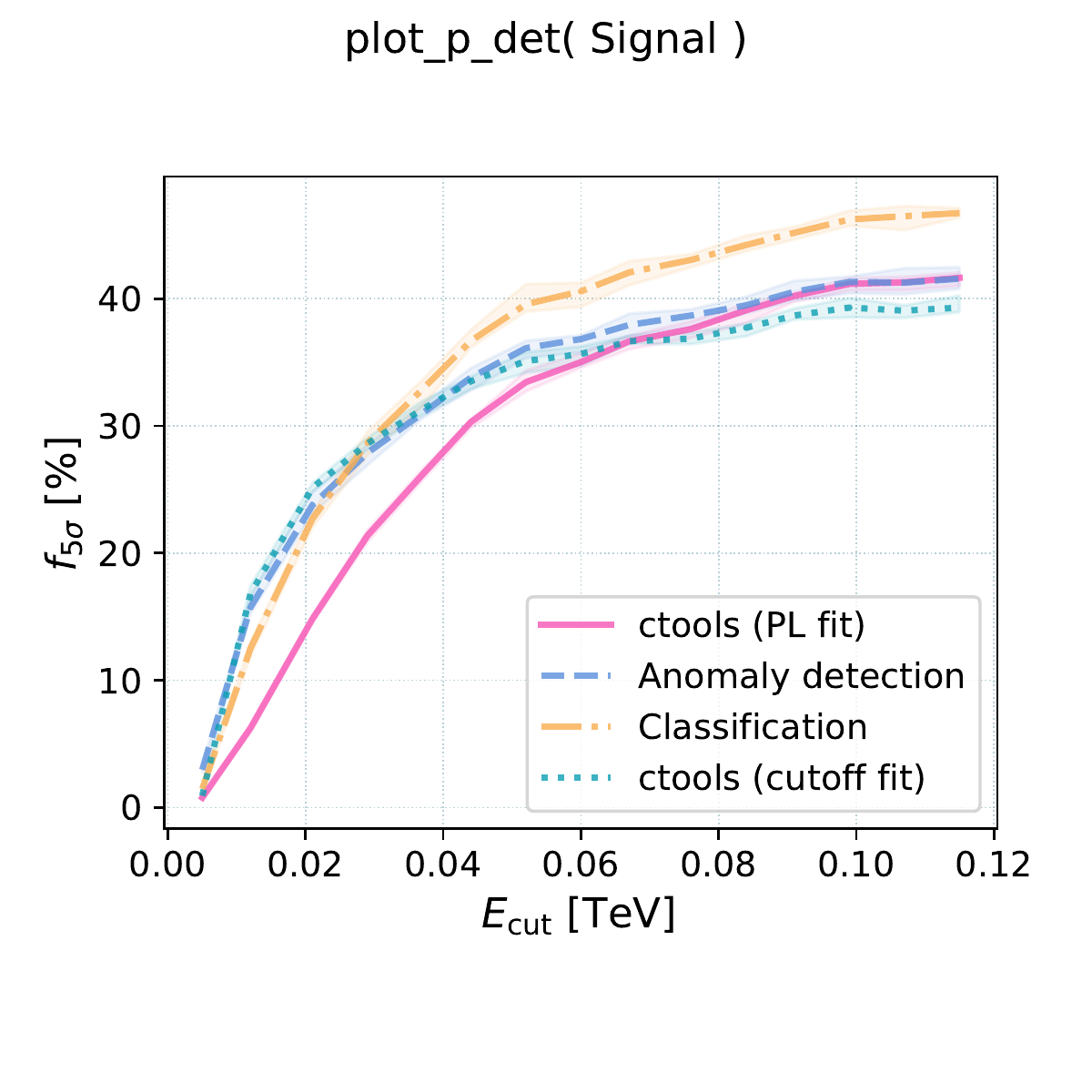}
                \put(40,193){\textbf{B}}
                \end{overpic}
                \end{center}
                \end{minipage}\hfill
                %
                %
                \begin{minipage}[c]{1\textwidth}
                \begin{center}
                \begin{minipage}[t]{1\textwidth}\begin{center}
                \caption{\label{FIGllgrbDetFrac} \textbf{Results
                  of the serendipitous \grb search analysis.}
                  \textbf{(A):}~The
                  probability to detect a \grb with at
                  least ${5\sigma}$ significance,
                  \pdet, derived with \andet, after accounting for trials.
                  The simulated sample includes different
                  combinations of redshift, $z$, and
                  isotropic equivalent luminosity, \liso,
                  spanning the expected properties of \llgrbs,
                  assuming \pl spectral models
                  (see \Autoref{EqGrbPlModelPL}).
                  \textbf{(B):}~Dependence of \pdet
                  on the cutoff energy, \ecut,
                  for bursts simulated as exponentially cutoff
                  \pls (see \Autoref{EqGrbPlModelCutoff}).
                  The shaded
                  regions correspond to~${1\sigma}$
                  statistical uncertainties on the values of \pdet.
                  Two alternative models are assumed
                  for detection with the likelihood method of \ctools,
                  an exponentially cutoff \pl,
                  and a simple \pl, as indicated. While the
                  \cls method was exclusively trained using simple \pl
                  examples, it effectively generalises and
                  identifies sources having cutoff spectra.}
                \end{center}\end{minipage}\hfill
                \end{center}
                \end{minipage}\hfill
                \vspace{5pt}
                \end{minipage}\hfill
            \end{figure*} 
            %
            Observations of \grbs at high energies are very interesting,
            \eg for the study of
            acceleration mechanisms in relativistic shocks.
            In many cases, \gamray emission extending
            up to \gev energies has been
            detected~\citep{Ackermann:2013zfa}.
            Recently, emission of up to hundreds of~\tev
            has been detected for the first time
            with ground-based Cherenkov Telescopes, which
            were following up alerts from other
            instruments~\citep{Arakawa:2019cfc, Acciari2019:455}.

            For the current example, we simulate an uninformed search
            scenario for the 
            upcoming Cherenkov Telescope Array (\cta; \cite{Acharya:2017ttl}).
            We focus on low-luminosity bursts
            (\llgrbs), a sub-class of long
            \grbs~\citep{Virgili:2008gp}, which
            have been connected
            to \sne having mildly relativistic outflows~\citep{Cano:2016ccp},
            and are potential sources of \ucrs and
            neutrinos~\citep{Murase:2013ffa, 2019ApJ...872..110B}.
            The true properties of \llgrbs are not
            well constrained by observations.
            Our sample therefore encompasses a wide
            range of spectral and temporal parameters.
            As such, it may be used to illustrate the detection
            prospects of such bursts, rather than to make precise
            predictions on observable rates.
            We correspondingly choose simple models,
            simulating the spectra of \llgrbs as
            either simple power laws (\pls) in
            energy and time,
            or as \pls having exponential cutoffs.

            We simulate observations for
            the Northern array of \cta
            using the \ctools
            analysis framework~\citep{2016A&A...593A...1K}.
            The simulations consists of \tit{\gamray-like} events.
            These correspond to true \gamrays, as well as
            to cosmic rays and electrons
            passing the nominal \cta selection and classification cuts.
            The inputs to the \rnn are event counts, ${n_{\gamma}}$,
            within ${\eta=\text{4}}$ logarithmically spaced energy bins
            between~30 and~200~\gev. They are
            integrated within circular
            regions of interest (\rois), over 1~\scnd periods
            (in order to probe the prompt emission phase
            of long \grbs).
            We construct the \rnn to represent ${\taurnn=\text{25}}$~\scnd
            of data. The first ${\tauenc=\text{20}}$~steps correspond
            to the background, and the final ${\taudec=\text{5}}$~steps to the 
            putative signal.
            In total, the \rnn 
            receives $\text{4}\times\text{25}=\text{100}$~inputs.

            For \andet, the \rnn is trained to predict
            \gamray-like counts in the absence of signals.
            In addition, we employ a \cls approach.
            Here, the network is trained with examples 
            of both background and
            signal events, where signals
            are injected over the ${\taudec=\text{5}}$~step interval.

            \Autoref{FIGllgrbDetFrac}~(A) shows the real-time
            discovery potential of the \rnn.
            As expected, bursts with lower redshifts and
            higher luminosities are more likely to be detected. In
            general, a large fraction of the parameter space is 
            accessible.
            In \Autoref{FIGllgrbDetFrac}~(B)
            we compare our algorithm with
            the likelihood-based method for source detection of \ctools.
            The \rnn performs similarly or better.
            For \andet, this is achieved
            without relying on
            instrument response functions.

            Contrary to \andet, \cls requires some assumption on
            sources as part of training. However,
            the performance of the
            method is shown to be robust to this constraint.
            In the current example, we simulate the 
            intrinsic spectra of \llgrbs as
            exponentially cutoff \pls.
            The classifier is
            trained exclusively with simple \pl examples. However,
            nonetheless it is able to generalise, and 
            outperforms the likelihood approach by achieving higher detection
            rates.
            We emphasise that for both \rnn configurations,
            training will primarily utilise real \cta data,
            once it becomes available,
            rather than simulations. 
            (For instance, for \cls a hybrid
            approach is possible, 
            where simulated signals are injected into
            background sequences from real data.)
            Correspondingly, the dependence on instrumental
            modelling is minimised using
            our framework.

            The advantage of the \rnn is 
            particularly evident when only \pl
            models are used as part of the \ctools likelihood fit.
            This represents a realistic strategy
            for blind searches, for which
            simple assumptions are generally made.
            Both our algorithms are comparatively agnostic
            to the properties of sources.
            They therefore
            enable real-time detection of a wider range of transients
            compared to standard techniques.
            This illustrates the merits of the methodology for
            unbiased searches, and the
            potential for unexpected discoveries.

    %
    \section{Discussion}

        Our method is optimised to minimise the
        bias on (real-time) detection of transients.
        It is distinguished from previous works,
        which have either incorporated explicit examples
        of signal events (mostly
        for classification), or have modelled
        the characteristics of backgrounds.
        Conversely, we use a data-driven strategy,
        which is less susceptible to the 
        pitfalls of unrepresentative training
        datasets.
        The predictions
        of the network are relatively robust against
        theoretical and systematic uncertainties
        on sources and instruments.
        Considering the simple architecture, searches
        may be conducted on different time scales.
        This may be done
        by modifying the value of \taurnn;
        by relating one or several \rnn steps to
        other temporal intervals; or by constructing
        (multiple) tests statistics 
        which span different scales. 
        In all cases the calibration phase 
        facilitates correct derivation of the final
        \pval-values for detection, accounting for trials.
        As illustrated for the \sne correlation study,
        the framework may also be used to derive limits
        from non-detections, though this requires
        instrument modelling.

        Our approach is relatively generalisable,
        enabling model-independent combination
        of observables.
        In principle, the framework facilitates
        sophisticated schemes of data fusion,
        where different data formats may be integrated
        consistently.
        However, in our nominal approach, this is not
        necessary. Subsequently, simple \rnn architectures may be used.
        This leads to robust predictions that
        do not depend on 
        extensive optimisation of hyper-parameters.

        In a realistic scenario of real-time searches,
        various systematic effects may initially be
        detected as transients.
        Experience with a particular instrument is 
        necessary in order to suppress these spurious signals. 
        Different strategies may be employed to this effect.
        For example:
        \begin{enumerate} [label=-,
        itemsep=1pt,topsep=0pt
        ]
            \item incorporating more sophisticated architectures, such as
            Bayesian networks~\citep{Shen:2019vep}, which may improve
            the calibration phase;
            \item using data that include these systematics for
            training (\eg the scrambling procedure
            for our neutrino point-source analysis);
            \item injecting known glitches as background events during training;
            \item correlating multiple observables that
            are susceptible to different systematics (\eg
            real-time combination of \gamrays and
            optical data), or cross-correlating \rois
            across the field of view;
            \item adding informative auxiliary observables to the \rnn;
            \item performing selection cuts pre-processing (\eg image
            quality cuts);
            \item performing filtering post-processing (\eg
            incorporating an additional network that
            would be trained either
            to classify systematic-induced events,
            or to perform \andet with regards to known 
            glitches; \cite{2018PhRvD..97j1501G});
            \item performing selection cuts post-processing, based
            on source modelling
            (\eg general constraints
            on time scales and energetics).
        %
        \end{enumerate}
        %
        In any case,
        even after accounting for systematics,
        true detections could correspond to
        different physical scenarios.
        In order to correctly characterise transients,
        \mwl/\mms follow-up will be essential.
        One of the primary applications of our method will be
        the effective identification of candidates for follow-up.

    %
    \acknowledgments
        We would like to thank the following
        people for numerous useful discussions:
        D.\,Biehl,
        D.\,Boncioli,
        Z.\,Bosnjak,
        A.\,Franckowiak,
        O.\,Gueta,
        T.\,Hassan,
        D.\,Horan,
        M.\,Krause,
        F.\,Longo,
        G.\,Maier,
        M.\,Nievas Rosillo,
        I.\,Oya,
        A.\,Palladino,
        E.\,Pueschel,
        R.\,R.\, Prado,
        L.\,Rauch,
        and
        W.\,Winter.
        This work has also gone through internal 
        review by the CTA Consortium.
        %


        We use the publicly available, all-sky point-source \IceCube data
        for years ${\text{2011--2012}}$~\citep{Aartsen:2016oji}.
        (See \href{http://doi.org/10.21234/B4F04V}{doi:10.21234/B4F04V}.)
        We also use publicly available data from the \antares experiment
        for the same time period~\citep{Adrian-Martinez:2014wzf}.
        (See \url{http://antares.in2p3.fr/publicdata2012.html}.)
        %
        We utilise simulations from the
        Photometric LSST Astronomical Time Series Classification Challenge
        (\plastic), an open data challenge to classify
        simulated astronomical time-series data~\citep{Kessler:2019qge}.
        (See \url{https://plasticc.org}; \href{http://doi.org/10.5281/zenodo.2539456}{doi:10.5281/zenodo.2535746}.) We also use \sncosmo, an
        open source library for supernova cosmology analysis~\citep{kyle_barbary_2016_168220}.
        (See~\href{https://sncosmo.readthedocs.io/en/v2.0.x/index.html}{doi:10.5281/zenodo.168220}.)
        This research makes use of \ctools~\citep{2016A&A...593A...1K}, a community-developed analysis package for Imaging Air Cherenkov Telescope data. \ctools is based on \gamlib, a community-developed toolbox for the high-level analysis of astronomical gamma-ray data. (See~\url{http://cta.irap.omp.eu/ctools/}; \url{http://cta.irap.omp.eu/gammalib/}.)
        We also use \cta-\irfs, provided by the \cta Consortium and Observatory (version \prodIrf). (See~\url{http://www.cta-observatory.org/science/cta-performance/}.)
        This research makes use of the open 
        source software, \tensorflow~\citep{tensorflow2015-whitepaper}.
        (See~\url{https://www.tensorflow.org/}.)

    \appendix
    \counterwithin{figure}{section}

    %
    \section{Statistical analysis} \label{sec:appendix:1}
        \subsec{Network architecture and inference pipeline} \label{sec:appendix:1:1}
            Our software pipeline and chosen \rnn
            architecture are shown in \Autoref{FIGannArch}.
            The \rnn is implemented
            using \tensorflow~\citep{tensorflow2015-whitepaper}.
            It may be decomposed into two
            elements, an \tit{encoder} and a
            \tit{decoder}~\citep{2014arXiv1406.1078C}.
            The \rnn accepts ${\taurnn = \tauenc + \taudec}$~time
            steps as input.
            The different steps are implemented as \lstm cells. A cell
            is composed of a pair of \lstm layers,
            respectively comprising~128 and~64
            hidden units (the set of parameters
            tuned during training).
            Each step receives a collection of
            (analysis-specific) ${\eta}$~inputs.
            The inputs are independently normalised, such that their
            nominal range of values for the background
            training sample is mapped to the interval,~${[0,1]}$.

            In our nominal approach we employ an
            \andet technique. 
            We utilise sequences of input data, $S(\taurnn, \eta)$, which
            for training correspond to the response of
            an instrument in the absence of signal events. 
            The \rnn
            is used to predict the
            expected background of the experiment, $B(\taudec, \eta)$,
            having the same data structure as $S(\taurnn, \eta)$.
            Transients are then detected as significant divergences
            from these predictions.
            Training involves minimising a \tit{loss function},
            which is defined as the \tit{mean squared error}
            between $B$ and $S$ for each set of~$\taurnn$--$\eta$.

            For \cls, an external layer
            is appended to the output decoder. 
            The latter maps $B(\taudec, \eta)$
            into \tit{logits}, ${\zeta(\taudec)}$.
            Each of these is a proxy for
            the \rnn probability density function (\pdf)
            for data of a given step to belong
            to the signal class~\citep{Goodfellow-et-al-2016}.
            Training proceeds by minimising a \tit{cross-entropy}
            loss function for the different steps, where
            the final probability is taken
            as the average,
            $\zetadec=\left< \zeta \right>_{\taudec}$.
            The corresponding \ts is based on the 
            ratio between the background and signal \pdfs
            for a given value of ${\zetadec}$~\citep{Cranmer:2015bka}.
            %
            As for \andet, calibration of test statistics
            into \pval-values is performed once
            following the training stage, using simulations.

\begin{figure*}[t]
                \begin{minipage}[c]{1\textwidth}

                \begin{minipage}[c]{1\textwidth}
                \begin{center}
                \includegraphics[trim=13mm 77mm 53mm 14mm,clip,width=1\textwidth]{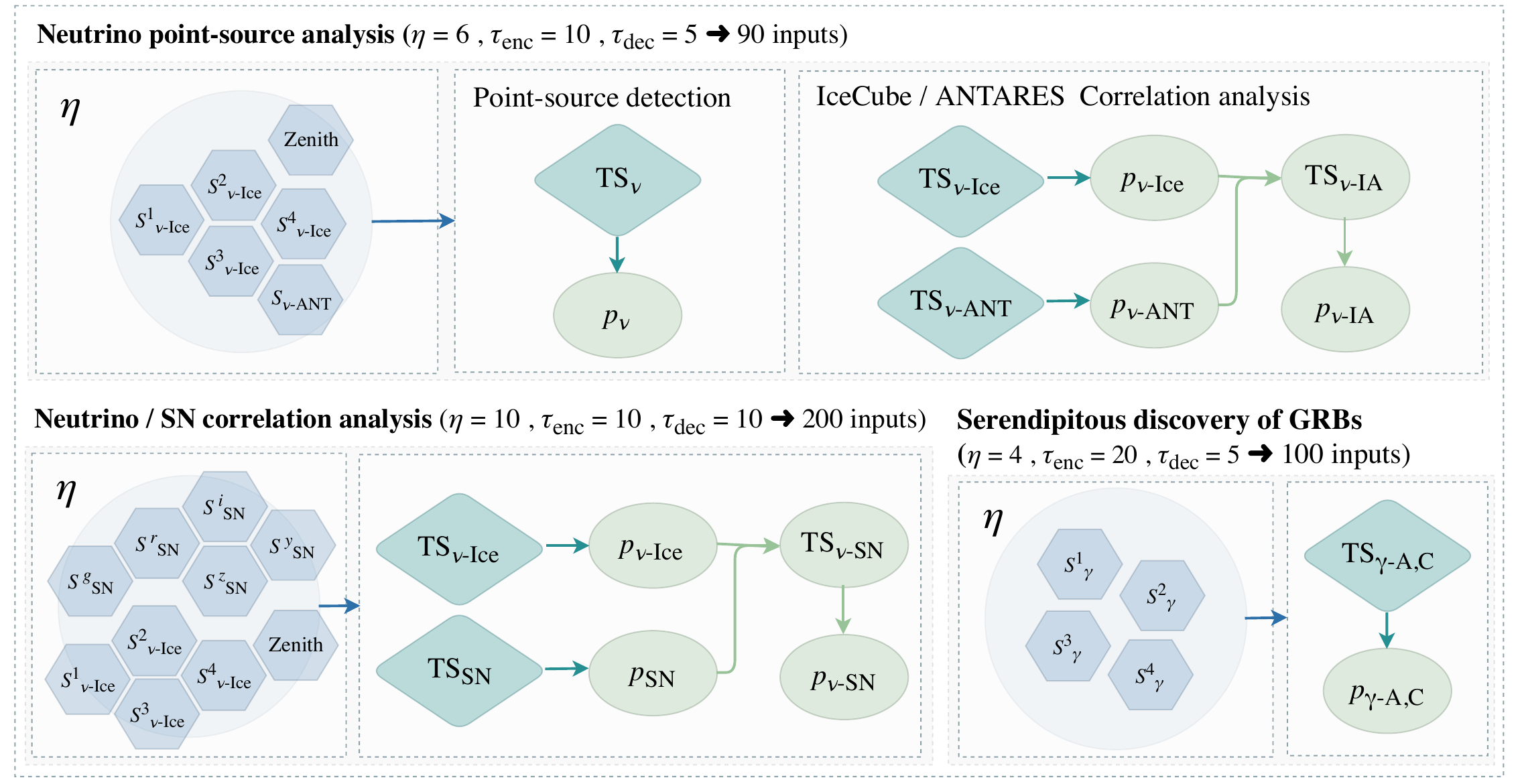}
                \end{center}
                \end{minipage}\hfill
                %
                %
                \begin{minipage}[c]{1\textwidth}
                \begin{center}
                \begin{minipage}[t]{1\textwidth}\begin{center}
                \caption{\label{FIGannPipeline}\textbf{Summary of
                  the specific
                  \rnn and pipeline configurations used
                  for the three analysis examples.}
                  The \rnn architectures are defined
                  by the number of encoder and decoder steps,
                  \tauenc and \taudec, and by the type and
                  number of inputs per
                  \rnn cell, $\eta$ (blue hexagons), as indicated.
                  The calibration
                  phases of the pipeline are defined by
                  the particular choices
                  of test statistics (\ts) and their
                  corresponding \pval-values for each analysis. 
                  The specific notation for each of the
                  analyses is defined in the text.}
                \end{center}\end{minipage}\hfill
                \end{center}
                \end{minipage}\hfill
                \vspace{15pt}
                \end{minipage}\hfill
            \end{figure*}

            In the following we detail the data reduction,
            source modelling, and definition of test statistics used for
            the example analyses presented in this study.
            A short summary is given in \Autoref{FIGannPipeline}.

        %
        \subsec{Neutrino point-source search} \label{sec:appendix:1:2}

            We search for clustering in 
            two publicly available all-sky
            $(\dec \in \{-85\dgr,\;85\dgr\})$
            neutrino samples.
            The first comprises \icecube event lists
            of track-like muon neutrino candidates, taken between
            ${\text{2011}}$ and ${\text{2012}}$ (MJD~55694--56415),
            with the ${\text{\textit{IC-86I}}}$ 
            detector configuration~\citep{Aartsen:2016oji}.
            The second is of \antares muon neutrinos, observed within
            the same time period~\citep{Adrian-Martinez:2014wzf}.
            Both samples include time-stamps, 
            reconstructed event directions, and angular uncertainties.
            In the case of \icecube,
            energy-proxy metrics and tabulated instrument
            response functions (\irfs) estimates
            are available as well.
            These public datasets are limited in scope.
            It is therefore not feasible to, \eg reliably
            correct for instrumental dead-time, or to estimate
            the efficiency of event reconstruction and selection cuts.
            An advantage of our methodology, is that such
            explicit corrections
            are not always necessary, so long
            as unbiased self-consistent observables
            can be derived.

            The nominal inputs to our \rnn are based on 
            \tit{neutrino event density} metrics with
            respect to a given location on the sky, ${\vec{x}}$.
            We define the density for a given neutrino as
            \begin{equation}
                s_{\nu}(\vec{x}) =
                \min{
                \left( 
                \frac{\delta_{\nu}}{\Delta_{\nu, \vec{x}}}
                \; , \;
                1
                \right)
                }
                \;.
                \label{EqNeutrinoEventDensityEstimates}
            \end{equation}
            Here ${\delta_{\nu}}$ is the angular uncertainty of 
            the event,
            and ${\Delta_{\nu, \vec{x}}}$ is the angular distance
            between the position of the neutrino and ${\vec{x}}$.
            We note that this definition of observables, while
            well motivated, is not unique.
            However, our framework is designed to provide self-consistent
            detection probabilities, given such informative input parameters.

            We integrate the event densities over 
            ${\text{24}}$~\hr time
            periods, which effectively avoids dependence of
            the event rates on \ra~\citep{Aartsen:2015wto}.
            For the \icecube sample, the data
            are split into four logarithmically
            spaced bins in the energy proxy, ${E_{\nu}}$,
            between ${\text{10}}$~\gev and ${\text{8}}$~\pev.
            For \antares the summation is inclusive in energy.
            We therefore have five density metrics as \rnn inputs,
            \begin{equation}
                \begin{split}
                S^{\epsilon}_{\nu\text{--Ice}}(\vec{x}) = &
                \sum_{
                \text{1~day}, \; \Delta E_{\nu}^{\epsilon}
                } 
                s_{\nu}(\vec{x})
                \;, \\
                \quad \text{and} \quad
                &
                S_{\nu\text{--ANT}} (\vec{x}) = 
                \sum_{
                \text{1~day}
                } 
                s_{\nu}(\vec{x})
                \;.
                \end{split}
                \label{EqNeutrinoEventDensitySum}
            \end{equation}
            with energy bins, $\epsilon\,\in\,\{1, 2, 3, 4\}$.

            The response of the \icecube and \antares
            detectors depends on the zenith
            of the observed event, which therefore is
            added as an auxiliary input to the \rnn. 
            We nominally use the \icecube events to determine
            the zenith for a particular sequence. Considering 
            the time of arrival and the location
            of the observatories, we calculate
            the corresponding zenith for the same putative
            source with 
            \antares. In cases where an \icecube source is not
            within the field of view of \antares,
            we use background data instead.
            In total, we have ${\eta=\text{6}}$~inputs per
            time step.
            The collection of inputs is derived for 
            ${\taurnn=\text{15}}$~days periods. 
            The first ${\tauenc=\text{10}}$~days are assumed to only
            contain background events. Transient
            signals are searched for within the next ${\taudec=\text{5}}$~days
            interval. In total, the \rnn 
            receives $\text{6}\times\text{15}=\text{90}$~inputs.

            We construct a background sample of
            ${\text{\powB{4}}}$ \taurnn sequences,
            where potential
            transient signals are removed
            by scrambling the events in \ra and 
            time of detection~\citep{Aartsen:2015wto}.
            The \rnn is trained to predict the
            neutrino event densities
            in each of the five days being probed, for a particular
            ${\vec{x}}$.
            The outputs of the network of the
            five event densities for a given day are denoted by 
            ${B^{\epsilon}_{\nu\text{--Ice}}}$ and
            ${B_{\nu\text{--ANT}}}$ for the two experiments.
            We use these to calculate a combined
            test statistic,
            \begin{equation}
                \text{TS}_{\nu}(\vec{x}) = 
                \text{TS}_{\nu\text{--Ice}}(\vec{x}) +
                \text{TS}_{\nu\text{--ANT}}(\vec{x})
                %
                \;,
                \label{EqNeutrinoTScombined}
            \end{equation}
            where
            \begin{equation}
                \begin{split}
                \text{TS}_{\nu\text{--Ice}} & (\vec{x}) =
                - \sum_{\text{5~days},\; \epsilon}
                \log_{10} \left(
                  \frac
                    {B^{\epsilon}_{\nu\text{--Ice}}}
                    {S^{\epsilon}_{\nu\text{--Ice}}}
                \right)
                \;, \\
                \quad \text{and}  & \quad
                \text{TS}_{\nu\text{--ANT}}(\vec{x}) = 
                - \sum_{\text{5~days}}
                \log_{10} \left(
                  \frac
                    {B_{\nu\text{--ANT}}}
                    {S_{\nu\text{--ANT}}}
                \right)
                \;.
                \end{split}
                \label{EqNeutrinoTSCorrelation}
            \end{equation}
            High values of ${\text{TS}_{\nu}}$ correspond
            to large discrepancies between (background) predictions
            and the corresponding true data.
            \ts-values are calibrated to \pval-values, ${p_{\nu}}$,
            as a function of the auxiliary parameter
            (the zenith, binned into~90
            intervals of~${\text{2}\dgr}$ width).

            Note that we take a generalisable approach, where
            the two datasets are combined on equal footing.
            More complicated schemes are also
            possible, \eg by introducing relative weights in
            the definition of ${\text{TS}_{\nu}}$.
            Similarly, different time-scales may be probed,
            either by modifying the definition of \taurnn,
            or that of ${\text{TS}_{\nu}}$.

            We also use the outputs of the \rnn to
            perform a correlation test between the
            \icecube and \antares events over the entire
            period of the dataset for each spatial position.
            For this purpose, we consider
            ${\text{TS}_{\nu\text{--Ice}}}$ and
            ${\text{TS}_{\nu\text{--ANT}}}$ individually.
            As part of the calibration stage,
            we derive the
            relation between \ts-values and \pval-values for
            each statistic independently. 
            This is done using
            simulations (as for ${\text{TS}_{\nu}}$ above).

            Our metrics for the correlation analysis are
            Pearson coefficients, based on the
            \pval-values of the two samples,
            ${p_{\nu\text{--Ice}}}$
            and ${p_{\nu\text{--ANT}}}$.
            Explicitly,
            \begin{equation}
                \begin{split}
                \text{TS}_{\nu\text{--IA}}(\vec{x}) = &
                \\ & \hspace{-35pt}
                \frac{
                \sum_{t}
                  \left( \rho^{t}_{\text{Ice}} - \left< \rho_{\text{Ice}} \right> \right)
                  \left( \rho^{t}_{\text{ANT}} - \left< \rho_{\text{ANT}} \right> \right)
                }{
                \sqrt{
                  \sum_{t}
                    \left( \rho^{t}_{\text{Ice}} - \left< \rho_{\text{Ice}} \right> \right)^{2}
                }
                \sqrt{
                  \sum_{t}
                    \left( \rho^{t}_{\text{ANT}} - \left< \rho_{\text{ANT}} \right> \right)^{2}
                }
                }
                \;,
                \end{split}
                \label{EqNeutrinoTSCorrelation}
            \end{equation}
            where we define
            $\rho^{t}_{\text{Ice}}(\vec{x}) \equiv - \log_{10} p_{\nu\text{--Ice}}$, and
            $\rho^{t}_{\text{ANT}}(\vec{x}) \equiv - \log_{10} p_{\nu\text{--ANT}}$.
            The summation is over the entire period
            of the dataset, with ${t}$ representing a particular
            \text{5}~days interval for a given sample.
            We proceed to derive the
            \pval-value for detection of a correlation signal,
            ${p_{\nu\text{--IA}}}$.
            For this purpose, we independently scramble 
            ${\rho^{t}_{\text{Ice}}}$ and
            ${\rho^{t}_{\text{ANT}}}$ in time
            and \ra\; This procedure is used
            to create multiple
            realisations of ${\text{TS}_{\nu\text{--IA}}(\vec{x})}$,
            for which the two samples are uncorrelated.
            Using these background distributions, 
            we account for spurious correlations, as well as
            for the number of spatial and temporal grid points.

        %
        \subsec{Correlation analysis between neutrinos \\and \ccsne} \label{sec:appendix:1:3}
            We simulate a neutrino stacking analysis.
            The general strategy is to identify an accumulated
            neutrino over-density 
            (based on ${\text{TS}_{\nu\text{-Ice}}}$)
            in spatio-temporal coincidence with 
            \ccsne~\citep{Senno:2017vtd}.
            %
            %
            \sne are simulated according to
            projected observations with the
            upcoming \lsst Wide-Fast-Deep
            survey~\citep{2009arXiv0912.0201L}.
            We utilise the public \plastic dataset,
            which was created as part of
            an open data challenge in preparation
            for \lsst~\citep{Kessler:2019qge}.
            The dataset represents the projected
            cadence and observing constraints for \lsst.
            For instance,
            it includes a prototype scheduler for
            science programme optimisation; realistic
            environmental conditions, such as
            weather and seeing; maintenance downtime;
            and instrumental artefacts.

            The inputs to the \rnn are of two types. 
            For a given time step, the
            first consists of the \icecube neutrino
            densities, ${S^{\epsilon}_{\nu\text{-Ice}}(\vec{x})}$.
            The second type consists of \lsst
            signal-to-noise metrics
            in several optical bands,
            $b\,\in\,\{g, r, i, z, y\}$.
            These are defined as
            \begin{equation}
                S_{\text{SN}}^{b}(\vec{x}) = 
                \frac{m_{b}}{\delta m_{b}}
                \;,
                \label{EqSNinputs}
            \end{equation}
            where ${m_{b}}$ and ${\delta m_{b}}$ respectively
            stand for an observed magnitude in band, $b$,
            and the corresponding uncertainty. As above,
            the zenith of observation is added as an auxiliary parameter.
            We thus have ${\eta=\text{10}}$~inputs per
            time step in total.
            We construct ${\taurnn=\text{20}}$~days data sequences, 
            with the first ${\tauenc=\text{10}}$~days representing
            the background, and the next
            ${\taudec=\text{10}}$~days the search period.
            In total the \rnn 
            receives $\text{10}\times\text{20}=\text{200}$~inputs.

            We begin by creating background samples. These represent
            periods of joint neutrino and optical observations,
            in which no transient exists in either messenger.
            The scrambling procedure described for the neutrino point-source
            analysis is used to derive ${S_{\nu\text{--Ice}}}$.
            We extract ${S_{\text{SN}}}$
            from simulation periods
            that lack transient signals, based on
            the true peak time of \sne light curves.
            The simulated \lsst observations include
            gaps in observations in some/all bands.
            We account for missing inputs by randomly interleaving
            background optical data in such gaps. 
            The background samples are used to train the \rnn,
            using ${\text{\powB{4}}}$ \taurnn sequences.

            For a particular sky position, we derive
            individual test statistics for source detection
            with each messenger. For neutrinos, we use 
            ${\text{TS}_{\nu\text{--Ice}}(\vec{x})}$ (here
            defined over ${\taudec=\text{10}}$~days intervals).
            For optical \sne, we have
            \begin{equation}
                \text{TS}_{\text{SN}}(\vec{x}) = 
                - \sum_{{\text{10~days},\; b}}
                \log_{10} \left(
                \frac
                {B^{b}_{\text{SN}}}
                {S^{b}_{\text{SN}}}
                \right)  
                \;,
                \label{EqNeutrinoTSsN}
            \end{equation}
            where ${B^{b}_{\text{SN}}}$ stands for the output of the network
            in a given band, $b$.
            In addition to high values of ${\text{TS}_{\text{SN}}}$,
            we impose a constraint on optical detections,
            that a \sn is observed over at least three nights in
            different bands.
            The correspondence between ${\text{TS}_{\nu\text{--Ice}}}$
            and ${\text{TS}_{\text{SN}}}$ and their respective
            \pval-values, ${p_{\nu\text{--Ice}}}$
            and ${p_{\text{SN}}}$, 
            is derived from simulations as a function
            of zenith.

            In the next step of the analysis, we simulate signal samples.
            The total number of simulated \sne up to a redshift
            of~${\text{0.06}}$ 
            is~${\Sim\text{2,000}~\yr^{-1}}$,
            following the star formation 
            rate~\citep{2012ApJ...753..152B}.
            Of these, about ${\text{57}\%}$ are detected
            with our \rnn with at least \stackSignifLow
            significance, and ${\text{42}\%}$ with \stackSignifHigh.
            We assume that the optically detected 
            \sne are in fact unobserved \grbs. 
            The \sne are generally characterised by
            the peak time of their light curves,
            which occurs on average 13~days after the 
            putative \grb~\citep{Cano:2016ccp}.
            As the exact time of the 
            emission is uncertain, we randomly 
            generate it as a Poisson
            process.
            We consider different values for \fjet, 
            the fraction of \sne for which the neutrino emission
            is directed towards the Earth.
            Depending on the physical model, the
            emission may \eg be collimated
            (relativistic jets; ${\fjet \ll1}$), or 
            quasi-spherical (shock breakouts; ${\fjet \Sim 1}$).
            We also account for misclassifications of \ccsne,
            by introducing a ${\text{10}\%}$ fake-rate,
            for which neutrino signals are not
            injected~\citep{Muthukrishna:2019wgc}.

            A neutrino flare is assumed to occur during the
            short prompt phase of the explosion.
            We relate the simulated flux of muon neutrinos to
            the energy deposited by \sne in \crays, \ecr.
            We use a flat \lcdm cosmology, with
            ${\Omega_{\mrm{m}} = 0.3}$ and the Hubble
            constant, ${\mrm{H}_{0} = 70~\mrm{km} \; \mrm{s}^{-1} \; \mrm{Mpc}^{-1}}$.
            The luminosity distance of optical \sne, ${D_{\mrm{L}}}$,
            is derived from their redshift.
            The fluence of simulated muon neutrinos is
            then given by
            $\neutF \propto \ecr / D^{2}_{\mrm{L}}$~\citep{Waxman:1998yy, Senno:2017vtd},
            assuming a \pl spectrum for the
            parent \crays with a spectral 
            index of~${-2}$, and a flare duration
            of ${\Sim\text{10}\text{--100}~\scnd}$.
            We derive the expected observed neutrino signal
            from the spectrum using the
            \irfs
            of \IceCube~\citep{Senno:2017vtd}.

            The test statistic of the stacked signal is defined as
            \begin{equation}
                \text{TS}_{\nu\text{--SN}}(\tau_{\text{LSST}}, \sigma_{\text{SN}}^{\text{min}}) = 
                \left<
                - \log_{10} p_{\nu\text{--Ice}}
                \right>
                %
                \;,
                \label{EqNeutrinoSnStackTs}
            \end{equation}
            where the average is over all \sne having 
            been detected during a survey of
            duration, ${\tau_{\text{LSST}}}$
            (either $\text{5}$ or $\text{10}$~\yr).
            In this context, detections are defined as
            events having an optical
            detection significance, ${\sigma_{\text{SN}}}$,
            higher than a given threshold, ${\sigma_{\text{SN}}^{\text{min}}}$
            (either \stackSignifLow or \stackSignifHigh).
            In order to derive the stacked detection \pval-value,
            ${p_{\nu\mrm{-SN}}}$, we simulate the background hypothesis;
            we impose the nominal optical 
            detection procedure, but do not
            inject any signals into the neutrino data.
            We generate~${\Sim\powB{7}}$ such
            realisations of a full \lsst survey of duration ${\tau_{\text{LSST}}}$, 
            accounting
            \eg for misidentified \sne, and for
            random coincidence between observables~\citep{Senno:2017vtd}.

        %
        \subsec{Serendipitous discovery of \grbs} \label{sec:appendix:1:4}

            Observations are simulated for
            the Northern array of \cta using \ctools~\citep{2016A&A...593A...1K},
            one of the proposed analysis frameworks for the observatory.
            The simulations produce \tit{\gamray-like} events.
            These correspond to true \gamrays, as well as
            to cosmic rays and electrons,
            which pass all selection and classification cuts.
            \ctools allows generation of background
            (\cray, electron) sky-maps, as well as
            of background$+$signal observations, by inclusion
            of source spectral models.
            We use \irfs optimised for 30~minutes
            observations at zenith
            angles of~${\text{20}\dgr}$.
            The \roi
            for the simulation is chosen as a circular
            area with a radius of ${\text{0.25}\dgr}$. It is centred 
            at the position of the putative source, and is
            displaced by ${\text{0.5}\dgr}$ from the centre
            of the camera.

            The inputs to the \rnn are \gamray-like event counts, ${n_{\gamma}}$,
            within ${\eta=\text{4}}$ logarithmically spaced energy bins, $E_{\gamma}$,
            between\;30 and\;200~\gev ($\epsilon\,\in\,\{1, 2, 3, 4\}$),
            integrated within 1~\scnd intervals within the \roi,
            \begin{equation}
                S^{\epsilon}_{\gamma} = 
                \sum_{
                \text{1~\scnd}, \; \Delta E_{\gamma}^{\epsilon}
                } 
                n_{\gamma}
                \;.
                \label{EqLLgrbEventCount}
            \end{equation}
            For \andet, we train the network with
            background-only \ctools simulations.
            The \rnn therefore predicts
            ${B^{\epsilon}_{\gamma}}$, the background
            event counts in each of the energy bins for each of the steps.
            We construct the \rnn to represent ${\taurnn=\text{25}}$~\scnd
            of data. The first ${\tauenc=\text{20}}$~steps correspond
            to the background, and the final ${\taudec=\text{5}}$~steps to the 
            putative signal.
            In total, the \rnn 
            receives $\text{4}\times\text{25}=\text{100}$~inputs.
            We train the network using ${\text{\powB{4}}}$
            background sequences.

            \begin{figure*}[tp]
                \begin{minipage}[c]{1\textwidth}
                \vspace{5pt}
                \begin{minipage}[c]{0.5\textwidth}
                \begin{center}
                \begin{overpic}[trim=0mm 12mm 0mm 13mm,clip,width=.98\textwidth]{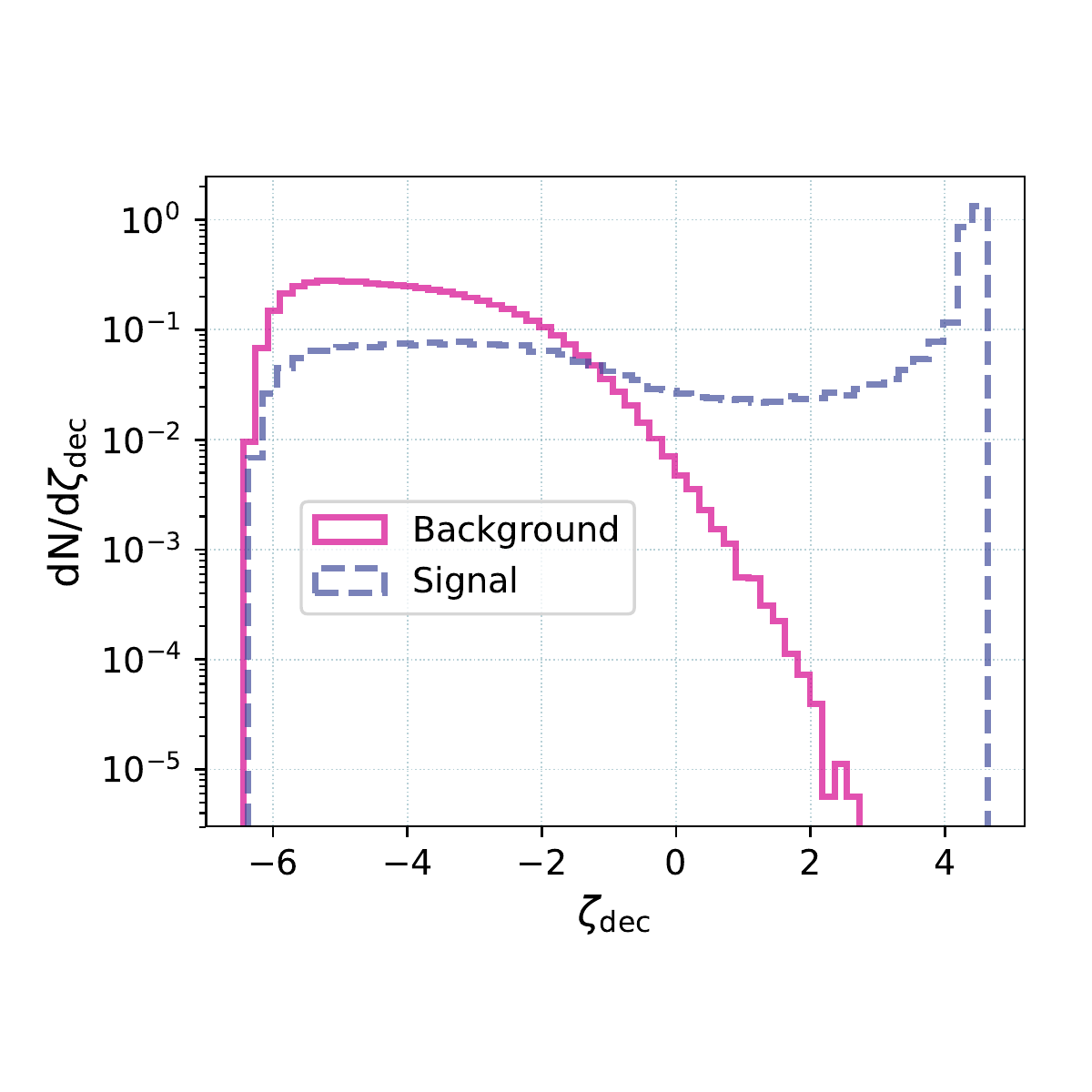}
                \put(50,193){\textbf{A}}
                \end{overpic}
                \end{center}
                \end{minipage}\hfill
                \begin{minipage}[c]{0.5\textwidth}
                \begin{center}
                \begin{overpic}[trim=0mm 12mm 0mm 13mm,clip,width=.98\textwidth]{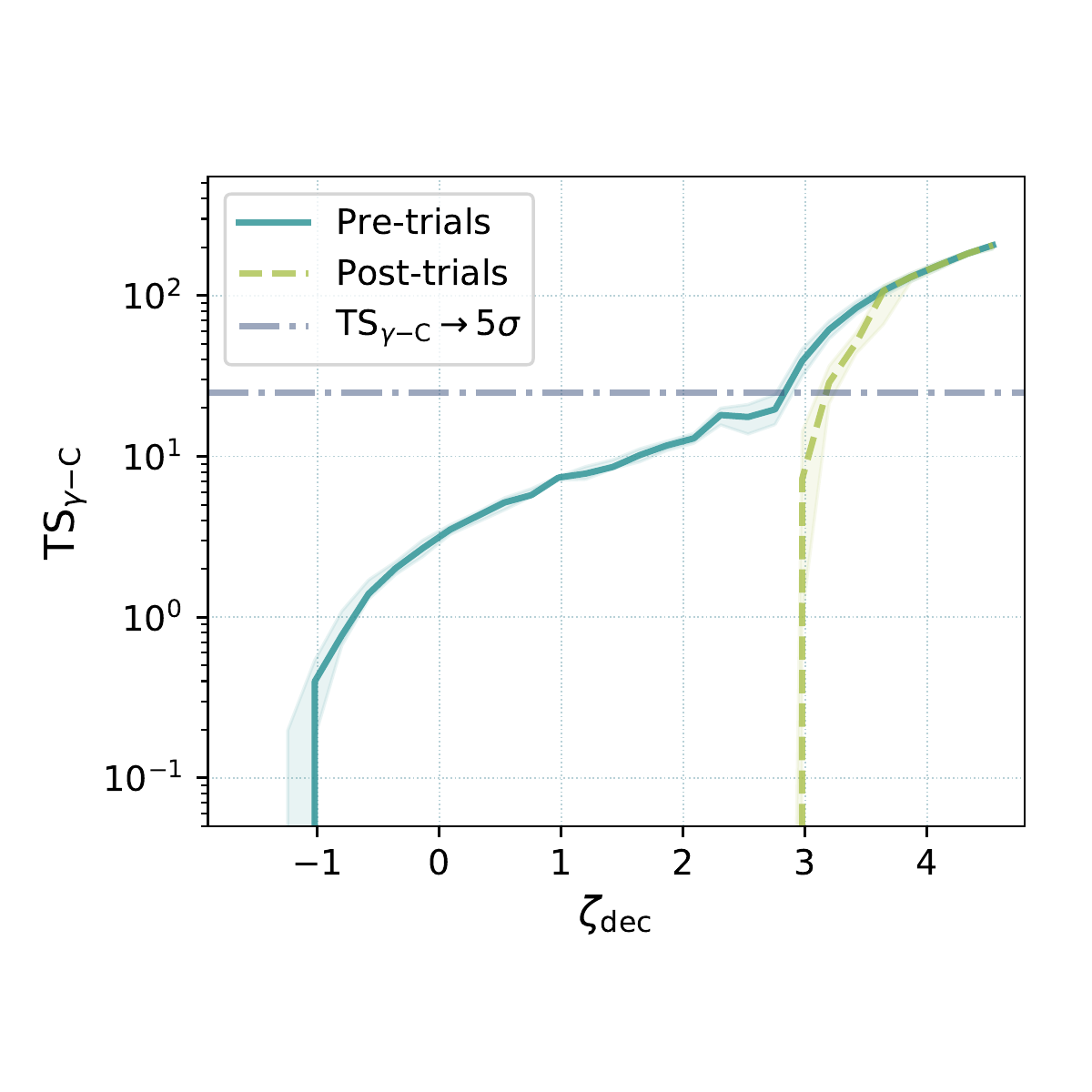}
                \put(50,193){\textbf{B}}
                \end{overpic}
                \end{center}
                \end{minipage}\hfill
                %
                %
                \begin{minipage}[c]{1\textwidth}
                \begin{center}
                \begin{minipage}[t]{1\textwidth}\begin{center}
                \caption{\label{FIGllgrbClasCalib} \textbf{Calibration of
                  the \cls output of the \rnn, \zetadec,
                  for the search for \llgrbs.}
                  \textbf{(A):}~The distributions of \zetadec
                  for the background and signal samples,
                  from which the probability density functions, 
                  ${\predClsB}$ and ${\predClsS}$,
                  are estimated.
                  \textbf{(B):}~Distributions of the test statistic
                  for classification, 
                  ${\text{TS}_{\gamma\text{--C}}}$, as a function
                  of \zetadec, before and after the correction for
                  trials, as indicated. 
                  The dashed--dotted horizontal line highlights
                  the value of ${\text{TS}_{\gamma\text{--C}}}$
                  corresponding to a ${\text{5}\sigma}$
                  detection threshold.}
                \end{center}\end{minipage}\hfill
                \end{center}
                \end{minipage}\hfill
                \vspace{5pt}
                \end{minipage}\hfill
            \end{figure*} 

            Our framework allows for explicit statistical models to be
            used, avoiding the need for calibration
            with background simulations.
            We illustrate this for the \llgrb analysis with Poissonian models
            of the signal and background$+$signal hypotheses.
            These are defined as 
            ${P_{\text{pois}}(k \vert \lambPois) = e^{-\lambPois}} \lambPois^{k} / k!$,
            given an observed number of events, $k$,
            and a rate,~\lambPois.
            We approximate \lambPois from the event count corresponding
            to a given hypothesis, integrating
            over the final ${\taudec=\text{5}}$~steps of the \rnn
            for a given energy bin,
            \begin{equation}
                \lambPois^{\epsilon}_{S} = 
                \sum_{
                \text{5~\scnd}
                } 
                S^{\epsilon}_{\gamma}
                \;, \quad \text{and} \quad
                \lambPois^{\epsilon}_{B} = 
                \sum_{
                \text{5~\scnd}
                } 
                B^{\epsilon}_{\gamma}
                \;.
                \label{EqLLgrbPoissRates}
            \end{equation}
            We then define the test statistic,
            \begin{equation}
                \text{TS}_{\gamma\text{--A}} = 
                - 2 \sum_{{\epsilon}}
                \ln \left(
                \frac
                {P_{\text{pois}}(S^{\epsilon}_{\gamma} \vert \lambPois^{\epsilon}_{B})}
                {P_{\text{pois}}(S^{\epsilon}_{\gamma} \vert \lambPois^{\epsilon}_{S})}
                \right)  
                \;.
                \label{EqLLgrbTS}
            \end{equation}
            We directly derive the pre-trials \pval-values for
            \andet, ${p_{\gamma\text{--A}}}$,
            using Wilks' theorem~\citep{Wilks:1938dza},
            which was validated using background simulations.

            In addition to \andet, we 
            employ a \cls approach. In this case, we require
            examples of signal events for training.
            To simulate the possible \gamray signatures of \llgrbs for \cta,
            we nominally model their spectra as
            simple power laws in
            energy, $E$, and time, $t$,
            \begin{equation}
            M_{\mrm{PL}}(E,t) \propto E^{-\Gamma} t^{-T}
            \;.
            \label{EqGrbPlModelPL}
            \end{equation}
            The spectral and temporal indices, $\Gamma$ and $T$,
            are parameters of the models.
            The true properties of these events are not well constrained,
            due to the scarcity of observations.
            We therefore consider a wide range of parameters.
            %
            Motivated by bright bursts, detected at high energies with
            \fermil,
            we randomise over likely values,
            $1.9 < \Gamma < 2.7$ and
            $0.8 < T < 2$~\citep{Ackermann:2013zfa}.
            The flux normalisation is randomly
            shifted with regards to these reference 
            events in redshift and luminosity
            to the expected ranges for \llgrbs~\citep{Acharya:2017ttl}.
            The observed spectra are corrected for interactions with the
            extragalactic background light (\ebl),
            which attenuates high-energy \gamrays~\citep{Franceschini:2008tp}.

            The network is trained for \cls using
            ${\text{\powB{4}}}$ background sequences
            and ${\text{\powB{4}}}$ signal sequences.
            Signals are injected over the ${\taudec=\text{5}}$~step interval,
            as part of the \ctools simulation.
            The network is trained with a wide range of such models
            having different flux normalisations, corresponding to
            different signal-to-noise ratios with
            respect to the background.
            The signal models incorporate \pl temporal decay.
            Correspondingly,
            late-time models are equivalent to early-time models with
            relatively lower flux normalisation.
            The inclusive composition of the training sample therefore 
            enhances generalisation (time-invariance) of the \rnn.

            While the inputs to the \rnn for \cls are the same as for \andet,
            the output in this case is a single number, $\zetadec$. The latter takes
            low values for input sequences that do not include transient signals,
            and high values in the presence of high-significance signals.
            We use the training sample to derive
            ${\predClsB}$ and ${\predClsS}$,
            the approximated \pdfs for
            the background and for the background$+$signal hypotheses
            (see \Autoref{FIGllgrbClasCalib}~(A)).
            Under the assumption that $\zetadec$ is
            monotonic with the ratio of \pdfs~\citep{Cranmer:2015bka},
            the \cls test statistic
            can be defined as
            \begin{equation}
                \text{TS}_{\gamma\text{--C}}(\zetadec) = 
                -2 \ln \left(
                \frac{\predClsB(\zetadec)}{\predClsB(\zetadec) + \predClsS(\zetadec)}
                \right)
                \;.
                \label{eqTsCls}
            \end{equation}
            We calibrate
            the relation between ${\text{TS}_{\gamma\text{--C}}}$
            and the corresponding pre-trials ${p_{\gamma\text{--C}}}$
            with background and signal simulations,
            using Wilks' theorem. The calibration
            is defined as a function of $\zetadec$,
            as illustrated in \Autoref{FIGllgrbClasCalib}~(B).

            In a realistic scenario, an uninformed search would be performed
            for \rois at different positions across 
            the field of view of \cta. The search
            would be repeated multiple times, depending on the
            definition of \taudec and on the amount of observing time.
            Correspondingly, the detection significance must be
            corrected for trials.
            We take,
            ${p_{\mrm{post}} = 1 - \left( 1 - p_{\mrm{pre}} \right)^{n_{\mrm{trial}}}}$,
            as the relation between pre- and
            post-trials probabilities.
            %
            We use the number of trials, 
            ${n_{\mrm{trial}}\Sim\powA{3.6}{7}}$, corresponding to
            searches over 100~\hr of observations, over
            the entire \cta field of view.

            It is currently difficult to estimate the rate of
            false detections for \cta, 
            as the observatory will explore
            a new regime of sensitivity~\citep{Acharya:2017ttl}.
            Experience with real data will improve
            the fidelity of detections. 
            As a first step, we performed systematic checks
            on the stability of 
            the \gamray-like event counts we use as input to the \rnn.
            The counts
            are susceptible to fluctuation due
            to imperfect \gamray reconstruction, and to
            uncertainties on the \irfs.
            In particular, energy dispersion
            below ${\Sim\text{50}~\gev}$ may result in
            migration between bins, and can
            change the energy threshold of the analysis.
            We studied these effects by
            comparing simulations where we vary the \irfs
            by their expected uncertainty
            (of up to $\text{10}\%$).
            We found that the propagated uncertainties
            on counts do not significantly affect our results.
            We also tested theoretical uncertainties on the
            effect of the \ebl.
            Comparing different models of the \ebl~\citep{Franceschini:2008tp, doi:10.1111/j.1365-2966.2010.17631.x, doi:10.1111/j.1365-2966.2012.20841.x},
            we found negligible impact on the observed spectra
            of \llgrbs. This is primarily due to the low redshift
            and energy regimes we consider in the current study.

            We also simulate a separate signal
            sample, where \llgrbs are modelled as
            \pls having exponential cutoffs,
            \begin{equation}
                M_{\mrm{EC}}(E,t) = M_{\mrm{PL}} \cdot \exp 
                \left(- \frac{E}{\ecut} \right)
                %
                \;.
                \label{EqGrbPlModelCutoff}
            \end{equation}
            In this case, we scan a range of cutoff energies,
            ${1 < \ecut < 120~\gev}$.
            The cutoff models are not used for training.
            Rather, we utilise them to illustrate the robustness
            of our methods (see \Autoref{FIGllgrbDetFrac}).
            As discussed above, the networks perform well
            for different source types.
            For instance, they enable identification of \grbs having
            cutoff spectral models, despite
            only being trained with simple \pl examples.

  \bibliography{bib}{}

\begin{thebibliography}{}
\expandafter\ifx\csname natexlab\endcsname\relax\def\natexlab#1{#1}\fi
\providecommand{\url}[1]{\href{#1}{#1}}
\providecommand{\dodoi}[1]{doi:~\href{http://doi.org/#1}{\nolinkurl{#1}}}
\providecommand{\doeprint}[1]{\href{http://ascl.net/#1}{\nolinkurl{http://ascl.net/#1}}}
\providecommand{\doarXiv}[1]{\href{https://arxiv.org/abs/#1}{\nolinkurl{https://arxiv.org/abs/#1}}}

\bibitem[{{Aartsen} {et~al.}(2013){Aartsen}, Abbasi, Abdou,
  {et~al.}}]{Aartsen:2013jdh}
{Aartsen}, M.~G., Abbasi, R., Abdou, Y., {et~al.} 2013, Science, 342, 1242856,
  \dodoi{10.1126/science.1242856}

\bibitem[{{Aartsen} {et~al.}(2017){Aartsen}, {Abraham}, {Ackermann},
  {et~al.}}]{Aartsen:2016oji}
{Aartsen}, M.~G., {Abraham}, K., {Ackermann}, M., {et~al.} 2017, Astrophys. J.,
  835, 151, \dodoi{10.3847/1538-4357/835/2/151}

\bibitem[{{Aartsen} {et~al.}(2015){Aartsen}, {Ackermann}, {Adams},
  {et~al.}}]{Aartsen:2015wto}
{Aartsen}, M.~G., {Ackermann}, M., {Adams}, J., {et~al.} 2015, Astrophys. J.,
  807, 46, \dodoi{10.1088/0004-637X/807/1/46}

\bibitem[{{Aartsen} {et~al.}(2018){Aartsen}, {Ackermann}, {Adams},
  {et~al.}}]{IceCube:2018cha}
---. 2018, Science, 361, 147, \dodoi{10.1126/science.aat2890}

\bibitem[{Abadi {et~al.}(2015)Abadi, Agarwal, Barham, Brevdo, Chen, Citro,
  Corrado, Davis, Dean, Devin, Ghemawat, Goodfellow, Harp, Irving, Isard, Jia,
  Jozefowicz, Kaiser, Kudlur, Levenberg, Man\'{e}, Monga, Moore, Murray, Olah,
  Schuster, Shlens, Steiner, Sutskever, Talwar, Tucker, Vanhoucke, Vasudevan,
  Vi\'{e}gas, Vinyals, Warden, Wattenberg, Wicke, Yu, \&
  Zheng}]{tensorflow2015-whitepaper}
Abadi, M., Agarwal, A., Barham, P., {et~al.} 2015.
\newblock \url{https://www.tensorflow.org/}

\bibitem[{{Abbott} {et~al.}(2017){Abbott}, {Abbott}, {Abbott}, {Acernese},
  {Ackley}, {Adams}, {Adams}, {Addesso}, {Adhikari}, {Adya}, \&
  et~al.}]{2017ApJ...848L..13A}
{Abbott}, B.~P., {Abbott}, R., {Abbott}, T.~D., {et~al.} 2017, \apjl, 848, L13,
  \dodoi{10.3847/2041-8213/aa920c}

\bibitem[{{Abdalla} {et~al.}(2019){Abdalla}, {Adam}, {Aharonian},
  {et~al.}}]{Arakawa:2019cfc}
{Abdalla}, H., {Adam}, R., {Aharonian}, F., {et~al.} 2019, Nature, 575, 464,
  \dodoi{10.1038/s41586-019-1743-9}

\bibitem[{Acciari {et~al.}(2019)Acciari, Ansoldi, Antonelli, Engels, Baack,
  Babic, Banerjee, Barres~de Almeida, Barrio, Becerra-Gonzalez, Bednarek,
  Bellizzi, Bernardini, Berti, Besenrieder, Bhattacharyya, Bigongiari, Biland,
  Blanch~Bigas, \& Nava}]{Acciari2019:455}
Acciari, V., Ansoldi, S., Antonelli, L.~A., {et~al.} 2019, Nature, 575, 455,
  \dodoi{10.1038/s41586-019-1750-x}

\bibitem[{{Acharya} {et~al.}(2017){Acharya}, {Agudo}, {Al Samarai},
  {et~al.}}]{Acharya:2017ttl}
{Acharya}, B.~S., {Agudo}, I., {Al Samarai}, I., {et~al.} 2017.
\newblock \doarXiv{1709.07997}

\bibitem[{{Ackermann} {et~al.}(2013){Ackermann}, {Ajello}, {Asano},
  {et~al.}}]{Ackermann:2013zfa}
{Ackermann}, M., {Ajello}, M., {Asano}, K., {et~al.} 2013, Astrophys. J.
  Suppl., 209, 11, \dodoi{10.1088/0067-0049/209/1/11}

\bibitem[{{Adrian-Martinez} {et~al.}(2014){Adrian-Martinez}, {Albert}, {Andre},
  {et~al.}}]{Adrian-Martinez:2014wzf}
{Adrian-Martinez}, S., {Albert}, A., {Andre}, M., {et~al.} 2014, Astrophys. J.,
  786, L5, \dodoi{10.1088/2041-8205/786/1/L5}

\bibitem[{Barbary {et~al.}(2016)Barbary, rbiswas4, Goldstein, Rodney, Jha,
  Wood-Vasey, Sofiatti, Feindt, Friesen, Barclay, Thomas, Craig, \&
  Barentsen}]{kyle_barbary_2016_168220}
Barbary, K., rbiswas4, Goldstein, D., {et~al.} 2016, sncosmo/sncosmo: v1.4.0,
  v1.4.0,  Zenodo, \dodoi{10.5281/zenodo.168220}

\bibitem[{{Bernstein} {et~al.}(2012){Bernstein}, {Kessler}, {Kuhlmann},
  {Biswas}, {Kovacs}, {Aldering}, {Crane}, {D'Andrea}, {Finley}, {Frieman},
  {Hufford}, {Jarvis}, {Kim}, {Marriner}, {Mukherjee}, {Nichol}, {Nugent},
  {Parkinson}, {Reis}, {Sako}, {Spinka}, \& {Sullivan}}]{2012ApJ...753..152B}
{Bernstein}, J.~P., {Kessler}, R., {Kuhlmann}, S., {et~al.} 2012, \apj, 753,
  152, \dodoi{10.1088/0004-637X/753/2/152}

\bibitem[{{Boncioli} {et~al.}(2019){Boncioli}, {Biehl}, \&
  {Winter}}]{2019ApJ...872..110B}
{Boncioli}, D., {Biehl}, D., \& {Winter}, W. 2019, \apj, 872, 110,
  \dodoi{10.3847/1538-4357/aafda7}

\bibitem[{Cano {et~al.}(2017)Cano, Wang, Dai, \& Wu}]{Cano:2016ccp}
Cano, Z., Wang, S.-Q., Dai, Z.-G., \& Wu, X.-F. 2017, Adv. Astron., 2017,
  8929054, \dodoi{10.1155/2017/8929054}

\bibitem[{Carleo {et~al.}(2019)Carleo, Cirac, Cranmer, Daudet, Schuld, Tishby,
  Vogt-Maranto, \& Zdeborová}]{Carleo:2019ptp}
Carleo, G., Cirac, I., Cranmer, K., {et~al.} 2019, Rev. Mod. Phys., 91, 045002,
  \dodoi{10.1103/RevModPhys.91.045002}

\bibitem[{{Cho} {et~al.}(2014){Cho}, {van Merrienboer}, {Gulcehre}, {Bahdanau},
  {Bougares}, {Schwenk}, \& {Bengio}}]{2014arXiv1406.1078C}
{Cho}, K., {van Merrienboer}, B., {Gulcehre}, C., {et~al.} 2014, arXiv
  e-prints, arXiv:1406.1078.
\newblock \doarXiv{1406.1078}

\bibitem[{{Choma} {et~al.}(2018){Choma}, {Monti}, {Gerhardt}, {Palczewski},
  {Ronaghi}, {Prabhat}, {Bhimji}, {Bronstein}, {Klein}, \&
  {Bruna}}]{2018arXiv180906166C}
{Choma}, N., {Monti}, F., {Gerhardt}, L., {et~al.} 2018, arXiv e-prints,
  arXiv:1809.06166.
\newblock \doarXiv{1809.06166}

\bibitem[{Cranmer {et~al.}(2015)Cranmer, Pavez, \& Louppe}]{Cranmer:2015bka}
Cranmer, K., Pavez, J., \& Louppe, G. 2015.
\newblock \doarXiv{1506.02169}

\bibitem[{Denton \& Tamborra(2018)}]{Denton:2017jwk}
Denton, P.~B., \& Tamborra, I. 2018, Astrophys. J., 855, 37,
  \dodoi{10.3847/1538-4357/aaab4a}

\bibitem[{Dominguez {et~al.}(2011)Dominguez, Primack, Rosario, Prada, Gilmore,
  Faber, Koo, Somerville, Pérez-Torres, Pérez-González, Huang, Davis,
  Guhathakurta, Barmby, Conselice, Lozano, Newman, \&
  Cooper}]{doi:10.1111/j.1365-2966.2010.17631.x}
Dominguez, A., Primack, J.~R., Rosario, D.~J., {et~al.} 2011, Monthly Notices
  of the Royal Astronomical Society, 410, 2556,
  \dodoi{10.1111/j.1365-2966.2010.17631.x}

\bibitem[{Franceschini {et~al.}(2008)Franceschini, Rodighiero, \&
  Vaccari}]{Franceschini:2008tp}
Franceschini, A., Rodighiero, G., \& Vaccari, M. 2008, Astron. Astrophys., 487,
  837, \dodoi{10.1051/0004-6361:200809691}

\bibitem[{Gebhard {et~al.}(2019)Gebhard, Kilbertus, Harry, \&
  Scholkopf}]{Gebhard:2019ldz}
Gebhard, T.~D., Kilbertus, N., Harry, I., \& Scholkopf, B. 2019, PhRvD, D100,
  063015, \dodoi{10.1103/PhysRevD.100.063015}

\bibitem[{George \& Huerta(2018{\natexlab{a}})}]{George:2016hay}
George, D., \& Huerta, E.~A. 2018{\natexlab{a}}, PhRvD, D97, 044039,
  \dodoi{10.1103/PhysRevD.97.044039}

\bibitem[{George \& Huerta(2018{\natexlab{b}})}]{George:2017pmj}
---. 2018{\natexlab{b}}, Phys. Lett., B778, 64,
  \dodoi{10.1016/j.physletb.2017.12.053}

\bibitem[{{George} {et~al.}(2018){George}, {Shen}, \&
  {Huerta}}]{2018PhRvD..97j1501G}
{George}, D., {Shen}, H., \& {Huerta}, E.~A. 2018, \prd, 97, 101501,
  \dodoi{10.1103/PhysRevD.97.101501}

\bibitem[{Gilmore {et~al.}(2012)Gilmore, Somerville, Primack, \&
  Dominguez}]{doi:10.1111/j.1365-2966.2012.20841.x}
Gilmore, R.~C., Somerville, R.~S., Primack, J.~R., \& Dominguez, A. 2012,
  Monthly Notices of the Royal Astronomical Society, 422, 3189,
  \dodoi{10.1111/j.1365-2966.2012.20841.x}

\bibitem[{Goodfellow {et~al.}(2016)Goodfellow, Bengio, \&
  Courville}]{Goodfellow-et-al-2016}
Goodfellow, I., Bengio, Y., \& Courville, A. 2016, Deep Learning (MIT Press)

\bibitem[{Graves \& Jaitly(2014)}]{Graves:2014:TES:3044805.3045089}
Graves, A., \& Jaitly, N. 2014, in Proceedings of the 31st International
  Conference on International Conference on Machine Learning - Volume 32,
  ICML'14 (JMLR.org), II--1764--II--1772.
\newblock \url{https://dl.acm.org/doi/10.5555/3044805.3045089}

\bibitem[{{Huerta} {et~al.}(2019){Huerta}, {Allen}, {Andreoni},
  {et~al.}}]{Huerta:2019rtg}
{Huerta}, E.~A., {Allen}, G., {Andreoni}, I., {et~al.} 2019, Nature Rev. Phys.,
  1, 600, \dodoi{10.1038/s42254-019-0097-4}

\bibitem[{{Kashiyama} {et~al.}(2013){Kashiyama}, {Murase}, {Horiuchi}, {Gao},
  \& {M{\'e}sz{\'a}ros}}]{2013ApJ...769L...6K}
{Kashiyama}, K., {Murase}, K., {Horiuchi}, S., {Gao}, S., \&
  {M{\'e}sz{\'a}ros}, P. 2013, \apj, 769, L6,
  \dodoi{10.1088/2041-8205/769/1/L6}

\bibitem[{{Kessler} {et~al.}(2019){Kessler}, {Narayan}, {Avelino},
  {et~al.}}]{Kessler:2019qge}
{Kessler}, R., {Narayan}, G., {Avelino}, A., {et~al.} 2019, PASP, 131, 094501,
  \dodoi{10.1088/1538-3873/ab26f1}

\bibitem[{Khan {et~al.}(2019)Khan, Huerta, Wang, Gruendl, Jennings, \&
  Zheng}]{Khan:2018opv}
Khan, A., Huerta, E.~A., Wang, S., {et~al.} 2019, Phys. Lett., B795, 248,
  \dodoi{10.1016/j.physletb.2019.06.009}

\bibitem[{{Kn{\"o}dlseder} {et~al.}(2016){Kn{\"o}dlseder}, {Mayer}, {Deil},
  {Cayrou}, {Owen}, {Kelley-Hoskins}, {Lu}, {Buehler}, {Forest}, \&
  {Louge}}]{2016A&A...593A...1K}
{Kn{\"o}dlseder}, J., {Mayer}, M., {Deil}, C., {et~al.} 2016, \aap, 593, A1,
  \dodoi{10.1051/0004-6361/201628822}

\bibitem[{LeCun {et~al.}(2015)LeCun, Bengio, \& Hinton}]{deepLearningReview}
LeCun, Y., Bengio, Y., \& Hinton, G. 2015, Nature, 521, 436 EP

\bibitem[{{LSST Science Collaboration} {et~al.}(2009){LSST Science
  Collaboration}, {Abell}, {Allison}, {Anderson}, {Andrew}, {Angel}, {Armus},
  {Arnett}, {Asztalos}, {Axelrod}, \& et~al.}]{2009arXiv0912.0201L}
{LSST Science Collaboration}, {Abell}, P.~A., {Allison}, J., {et~al.} 2009,
  arXiv e-prints.
\newblock \doarXiv{0912.0201}

\bibitem[{Malhotra {et~al.}(2015)Malhotra, Vig, Shroff, \&
  Agarwal}]{Malhotra2015LongST}
Malhotra, P., Vig, L., Shroff, G., \& Agarwal, P. 2015, in 23rd European
  Symposium on Artificial Neural Networks (ESANN), Bruges, Belgium.
\newblock \url{http://www.i6doc.com/en/}

\bibitem[{Meszaros {et~al.}(2019)Meszaros, Fox, Hanna, \&
  Murase}]{Meszaros:2019xej}
Meszaros, P., Fox, D.~B., Hanna, C., \& Murase, K. 2019, Nature Rev. Phys., 1,
  585, \dodoi{10.1038/s42254-019-0101-z}

\bibitem[{{Morgan}(2019)}]{2019ICRC...36..963M}
{Morgan}, R. 2019, in International Cosmic Ray Conference, Vol.~36, 36th
  International Cosmic Ray Conference (ICRC2019), 963.
\newblock \doarXiv{1907.07193}

\bibitem[{Murase \& Ioka(2013)}]{Murase:2013ffa}
Murase, K., \& Ioka, K. 2013, PhRvL, 111, 121102,
  \dodoi{10.1103/PhysRevLett.111.121102}

\bibitem[{Muthukrishna {et~al.}(2019)Muthukrishna, Narayan, Mandel, Biswas, \&
  Hložek}]{Muthukrishna:2019wgc}
Muthukrishna, D., Narayan, G., Mandel, K.~S., Biswas, R., \& Hložek, R. 2019,
  PASP, 131, 118002, \dodoi{10.1088/1538-3873/ab1609}

\bibitem[{Pimentel {et~al.}(2014)Pimentel, Clifton, Clifton, \&
  Tarassenko}]{PIMENTEL2014215}
Pimentel, M.~A., Clifton, D.~A., Clifton, L., \& Tarassenko, L. 2014, Signal
  Processing, 99, 215 , \dodoi{https://doi.org/10.1016/j.sigpro.2013.12.026}

\bibitem[{{Pruzhinskaya} {et~al.}(2019){Pruzhinskaya}, {Malanchev}, {Kornilov},
  {Ishida}, {Mondon}, {Volnova}, \& {Korolev}}]{2019MNRAS.489.3591P}
{Pruzhinskaya}, M.~V., {Malanchev}, K.~L., {Kornilov}, M.~V., {et~al.} 2019,
  \mnras, 489, 3591, \dodoi{10.1093/mnras/stz2362}

\bibitem[{Senno {et~al.}(2018)Senno, Murase, \& Meszaros}]{Senno:2017vtd}
Senno, N., Murase, K., \& Meszaros, P. 2018, JCAP, 1801, 025,
  \dodoi{10.1088/1475-7516/2018/01/025}

\bibitem[{Shen {et~al.}(2019)Shen, Huerta, Zhao, Jennings, \&
  Sharma}]{Shen:2019vep}
Shen, H., Huerta, E.~A., Zhao, Z., Jennings, E., \& Sharma, H. 2019.
\newblock \doarXiv{1903.01998}

\bibitem[{Sutskever {et~al.}(2014)Sutskever, Vinyals, \&
  Le}]{Sutskever:2014:SSL:2969033.2969173}
Sutskever, I., Vinyals, O., \& Le, Q.~V. 2014, in Proceedings of the 27th
  International Conference on Neural Information Processing Systems - Volume 2,
  NIPS’14 (Cambridge, MA, USA: MIT Press), 3104–3112

\bibitem[{Virgili {et~al.}(2009)Virgili, Liang, \& Zhang}]{Virgili:2008gp}
Virgili, F., Liang, E., \& Zhang, B. 2009, MNRAS, 392, 91,
  \dodoi{10.1111/j.1365-2966.2008.14063.x}

\bibitem[{Waxman \& Bahcall(1999)}]{Waxman:1998yy}
Waxman, E., \& Bahcall, J.~N. 1999, PhRvD, D59, 023002,
  \dodoi{10.1103/PhysRevD.59.023002}

\bibitem[{{Wei} \& {Huerta}(2020)}]{Wei:2019zlc}
{Wei}, W., \& {Huerta}, E.~A. 2020, Physics Letters B, 800, 135081,
  \dodoi{10.1016/j.physletb.2019.135081}

\bibitem[{Wilks(1938)}]{Wilks:1938dza}
Wilks, S.~S. 1938, Annals Math. Statist., 9, 60,
  \dodoi{10.1214/aoms/1177732360}

\bibitem[{{Williamson} {et~al.}(2019){Williamson}, {Modjaz}, \&
  {Bianco}}]{2019ApJ...880L..22W}
{Williamson}, M., {Modjaz}, M., \& {Bianco}, F.~B. 2019, \apjl, 880, L22,
  \dodoi{10.3847/2041-8213/ab2edb}

\bibitem[{{Zackay} \& {Ofek}(2017)}]{2017ApJ...836..187Z}
{Zackay}, B., \& {Ofek}, E.~O. 2017, \apj, 836, 187,
  \dodoi{10.3847/1538-4357/836/2/187}

\bibitem[{{Zevin} {et~al.}(2017){Zevin}, {Coughlin}, {Bahaadini},
  {et~al.}}]{Zevin:2016qwy}
{Zevin}, M., {Coughlin}, S., {Bahaadini}, S., {et~al.} 2017, CQGra, 34, 064003,
  \dodoi{10.1088/1361-6382/aa5cea}

\end{thebibliography}
  \bibliographystyle{aasjournal}

\end{document}